\documentclass[12pt]{article}
\usepackage{amssymb,tabularx,multirow,amsmath,natbib, epsfig, threeparttable, amstext}
\usepackage{setspace}
\usepackage{verbatim}
\usepackage{times}
\usepackage{media9} 
\usepackage{floatrow}
\usepackage{float}
\usepackage{framed}
\usepackage{caption}
\usepackage{subcaption}
\usepackage{floatrow}
\usepackage{framed}
\usepackage{bbm}
\usepackage{algorithm} 
\usepackage{algorithmic}  
\usepackage{hyperref}
\usepackage[algo2e]{algorithm2e} 
\usepackage{framed}
\usepackage{varwidth}

\newcommand{\R}{\mathbb{R}}

\newcommand{\mbf}[1]{\mathbf{#1}}
\newcommand{\mbv}[1]{\mbox{\boldmath$#1$\unboldmath}}

\def\bh{\mathbf{h}}
\def\bg{\mathbf{g}}

\def\bA{\mathbf{A}}

\def\bI{\mathbf{I}}

\def\bV{\mathbf{V}}
\def\bU{\mathbf{U}}
\def\bW{\mathbf{W}}
\def\bX{\mathbf{X}}
\def\bY{\mathbf{Y}}
\def\bZ{\mathbf{Z}}

\begin{document}
\doublespacing
\title{\bf  Bayesian Recurrent Neural Network Models for Forecasting and Quantifying Uncertainty in Spatial-Temporal Data}
\author{Patrick L. McDermott\thanks{Department of Statistics, University of Missouri, 146 Middlebush Hall,
Columbia, MO 65211 USA; E-mail: plmyt7@mail.missouri.edu (corresponding author)} \hspace{5mm} Christopher K.
Wikle \\ Department of Statistics \\ University of Missouri}
 \maketitle
 
\begin{abstract}

Recurrent neural networks (RNNs) are nonlinear dynamical models commonly used in the machine learning and dynamical systems literature to represent complex dynamical or sequential relationships between variables. More recently, as deep learning models have become more common, RNNs have been used to forecast increasingly complicated systems. Dynamical spatio-temporal processes represent a class of complex systems that can potentially benefit from these types of models. Although the RNN literature is expansive and highly developed, uncertainty quantification is often ignored.  Even when considered, the uncertainty is generally quantified without the use of a rigorous framework, such as a fully Bayesian setting. Here we attempt to quantify uncertainty in a more formal framework while maintaining the forecast accuracy that makes these models appealing, by presenting a Bayesian RNN model for nonlinear spatio-temporal forecasting. Additionally, we make simple modifications to the basic RNN to help accommodate the unique nature of nonlinear spatio-temporal data. The proposed model is applied to a Lorenz simulation and two real-world nonlinear spatio-temporal forecasting applications 

 \end{abstract}

\textbf{Keywords:}
recurrent neural network, Bayesian machine learning, nonlinear dynamical models, long-lead forecasting, spatial-temporal

\section{Introduction}\label{sec:Intro}
Nonlinear and quasilinear spatio-temporal data can be found throughout the engineering, biological, geophysical and social sciences. Some examples of such processes include animal or robotic interactions, local economic forecasting, river flow forecasting, visual motion capture, and radar precipitation reflectivity nowcasting, to name a few. The nonlinearity in these systems makes forecasting and quantifying uncertainty difficult from both a modeling and computational perspective. While statistical forecasting of univariate nonlinear time-series processes is relatively well-developed \citep[][]{FanANDYao,billings2013nonlinear}, nonlinear multivariate systems have seen much less progress in statistics. Dynamical spatial-temporal models (DSTMs) are multivariate systems that have the added challenge of characterizing interactions between different scales of variability while simultaneously facing the curse-of-dimensionality that is exacerbated for nonlinear parametric spatio-temporal models \citep[e.g.,][]{wikle2015modern}. Both of these issues, along with a need for flexibility, can lead to intense computational demands for nonlinear DSTMs and nonlinear multivariate processes.

Some more recent nonlinear statistical DSTMs include threshold or regime switching models \citep[e.g.,][]{berliner2000long,wu2013hierarchical}, agent (individual)-based models \citep[e.g.,][]{hooten2010statistical}, general quadratic nonlinear (GQN) models \citep[][]{wikle2010general}, analog models \citep[][]{mcdermott2016model}, and mechanistic nonlinear models \citep[][]{richardson2017sparsity}. While such models have shown success for particular systems, more flexible models are often needed for highly nonlinear systems with complex latent relationships. Furthermore, with only a few exceptions, it can be quit difficult to explicitly specify the nonlinearities in these systems. One exception includes using physically motivated models such as stochastic partial differential equations \citep[e.g.,][]{wikle2010general,CandW2011,richardson2017sparsity}, although this requires some {\it a priori} knowledge of the dynamics in the system. Due to the many challenges associated with modeling nonlinear spatio-temporal processes, much of the statistical development of these models has lagged behind other disciplines such as dynamical systems and machine learning.

One of the many appealing aspects of machine learning methods is their ability to extract salient features and relationships from complex high-dimensional data, particularly for forecasting and classification. Spatio-temporal processes are a strong candidate for machine learning methods due to the complex interactions and high-dimensionality that are ubiquitous in these processes. While there have been past attempts to apply machine learning methods, such as feed-forward neural networks \citep[e.g.,][]{tang2000skill} and deep learning models \cite[e.g.,][]{dixon2017deep} to nonlinear spatial-temporal processes, the explicit accounting for dynamics in these processes has been less of a focus. Moreover, although feed-forward neural networks provide a convenient framework for modeling multivariate processes, they are not designed to explicitly capture time-sequential dynamical interactions between variables.  As often noted in the dynamical systems literature, explicitly modeling the dynamics is often paramount to successfully forecasting such systems. Recurrent neural networks (RNNs) represent a machine learning model with the potential to effectively model the nonlinear dynamics in multivariate sequential systems such as spatio-temporal processes.

First popularized in the 1980s, RNNs fell out of favor, in part, because of the so-called vanishing gradient problem that makes these models extremely difficult to estimate with back-propagation. More recently, as deep learning models have gained in popularity, solutions such as the gradient normal clipping strategy \cite[][]{pascanu2013difficulty} have eased the overall implementation burden of RNNs. As RNNs have become more manageable from an estimation perspective, they have increasingly been used to model complicated sequential forecasting problems such as visual object tracking \citep[][]{ning2016spatially}, speech recognition \citep[][]{yildiz2013birdsong}, and text generation \citep[][]{graves2013generating}, just to name a few. Simultaneously, RNNs have also seen a rise in usage in the dynamic systems literature due to their ability to replicate complex attractor dynamics that  are often present in chaotic systems \citep[][]{jaeger2001echo}. Thus, RNNs provide a black-box method that can capture dynamical relationships for problems where it is either difficult to specify these relationships {\it a priori} or little information is available on the specific form of these relationships. Importantly, RNNs fill this void by providing a mechanism for capturing complex sequential relationships between variables, thus providing a modeling tool for a broad set of dynamical problems. 

As RNNs have become more prevalent, a variant of the original RNN model, referred to as an {\it echo-state network} (ESN) \citep[][]{lukovsevivcius2009reservoir} has become a staple in the dynamical systems literature for solving nonlinear forecasting problems. ESNs are extremely appealing because they retain much of the forecast accuracy of a RNN at a fraction of the computational cost. In essence, ESNs simulate randomly the parameters that make up the hidden states of a RNN (see below), thus reducing the problem to a traditional regression type problem. Although the methodology described here is more closely related to the RNN framework than the ESN, we do borrow and discuss ideas from the ESN literature to motivate choices pertaining to the proposed model. For a spatial-temporal example of an ESN model see \cite{mcdermott2017ensemble}. 

Despite the broad size and overall scope of the RNN literature, these models are almost always presented without considering uncertainty. The few attempts at quantifying uncertainty are generally presented in an {\it ad hoc} fashion, without a formal probability based framework. Conversely, as previously discussed, the literature on statistical modeling of nonlinear dynamical spatio-temporal systems does consider uncertainty quantification but is not well-developed, especially compared to its linear counterpart. We address both of these issues by proposing a Bayesian spatial-temporal RNN model in which the forecasting strength of a traditional RNN is preserved, while also producing comprehensive uncertainty measures. In particular, we introduce a RNN model within a fully Bayesian framework that accounts for uncertainty in both parameters and data in a rigorous fashion.

While others have used Bayesian modeling within the RNN framework \citep[e.g.,][]{chatzis2015sparse,chien2016bayesian,gan2016scalable}, to our knowledge this is the first fully Bayesian RNN trained with traditional Markov Chain Monte Carlo (MCMC) methods. By using MCMC methods, the proposed model and algorithm can more accurately measure uncertainty compared to traditional optimization methods or variational Bayesian methods. Recently, Bayesian methods such as stochastic gradient MCMC (SG-MCMC) have shown promise as an estimation tool for high-dimensional RNNs \citep[i.e.,][]{gan2016scalable}. These stochastic gradient algorithms typically require the partitioning of the data to create so-called mini-batches. Spatio-temporal models often involve explicit dependencies between data points in space and/or time along with hierarchical relationships. Therefore, it may be difficult or impossible to partition the data in this way, which may make such stochastic algorithms prohibitive for some spatial-temporal problems.

We introduce multiple extensions to the traditional RNN model at both the data and latent stage of the model, with the dual aim of facilitating estimation and improving the forecasting ability of the model. The proposed extensions incorporate mechanisms from both the ESN and dynamical systems literature. Furthermore, we regularize the parameters in the model by proposing priors that help mitigate over-parameterization, inspired by traditional ESN models. Similar to traditional RNNs, fitting a RNN within a fully Bayesian framework presents a multitude of computational issues. To assist with computation, we propose using dimension reduction to deal with high-dimensional spatial-temporal processes. In addition, within a MCMC paradigm, we borrow the idea of including expansion parameters in the model from the data augmentation literature \citep[e.g.,][]{liu1999parameter,hobert2008theoretical,hobert2011data}, to assist with sampling the highly dependent parameters that make up a RNN.

Section 2 describes the proposed Bayesian spatio-temporal RNN model, along with various modeling details. Next, Section 3 goes through the specifics of the MCMC algorithm developed to implement the model. In the beginning of Section 4 the choices made for the setup of the model are described in detail.  Section 4 continues with a simulated multiscale Lorenz dynamical system example, followed by a long-lead sea surface temperature (SST) forecasting problem and a United States (U.S.) state-level unemployment rate application. Finally, we end with a concluding discussion on the approach, along with possible future extensions in Section 5.

\section{Spatio-Temporal Recurrent Neural Network}
\subsection{Traditional Recurrent Neural Network}

Suppose we are interested in the $n_y$-dimensional spatial-temporal response vector $\bY_t$ at time $t$ with corresponding input vector $\bX_t$ of dimension $n_x$, with a one being the first element of $\bX_t$ corresponding to an intercept term (or  bias term). Then, the traditional RNN model for multivariate data \citep[e.g.,][]{chung2015gated} is defined as follows for $t=1,\dots,T$:
\begin{eqnarray}
\mbox{data stage: }  & \; & {\mbf Y}_t = g(\bV\bh_t),  \label{eq:datastage} \\
\mbox{hidden stage: } & \; & \bh_t = f(\bW \bh_{t-1} + \bU \bX_t), \label{eq:h1}
\end{eqnarray}
where $\bh_t$ is a $n_h$-dimensional vector of hidden state variables, $\bW$ is a square $n_h \times n_h$ weight matrix, $\bU$ is a $n_h \times n_x$ weight matrix, and $\bV$ is a $n_y \times n_h$ weight matrix. The function $g(\cdot)$ is an activation function that creates a mapping between the response and the hidden states, and $f(\cdot)$ denotes the activation function for the hidden layer. For a continuous response vector, $g(\cdot)$ is simply the identity function, although this setup can also handle categorical data by allowing $g(\cdot)$ to be the softmax function. Nonlinearity is induced in the RNN model through the form of $f(\cdot)$, which is typically defined to be the hyperbolic tangent function (as is assumed throughout this article).

Furthermore, the square weight matrix $\bW$ can be thought of analogously to a transition matrix in a typical vector autoregressive (VAR) model. That is, $\bW$ models the latent dynamic connections between the various hidden states. Thus, underlying nonlinear interactions between variables or locations can effectively be modeled within this framework through $\bW$. Having a mechanism to capture these interactions is often vital when modeling nonlinear spatio-temporal processes \citep[e.g.,][]{wikle2015modern}. Critically, the hidden states extract and supply salient hidden dynamic features from the data. Ideally, the hidden states will represent a general set of dynamical patterns from the input data, thus allowing the $\bV$ parameters to appropriately weight these patterns. While the RNN defined in (\ref{eq:datastage}) and (\ref{eq:h1}) has shown success at forecasting a variety of different systems, the model lacks any explicit error terms, and thus, does not contain a mechanism to formally account for uncertainty in the data, model, or parameters.

\subsection{Bayesian Spatio-Temporal Recurrent Neural Network}
In this section we introduce the Bayesian spatio-temporal RNN, referred to hereafter as the BAST-RNN model. Borrowing the notation introduced in the previous section, the BAST-RNN model is defined as follows:
\begin{eqnarray}
\mbox{data stage: }  & \; & {\mbf Y}_t ={\mbv \mu}+  \bV_1\bh_t + \bV_2\bh_t^2 + {\mbv \epsilon}_t, \hspace{.3cm} \;\; {\mbv \epsilon}_t \; \sim \; \text{Gau}({\mbf 0},{\mbf R}_t),   \label{eq:datastage2} \\
\mbox{hidden stage: } & \; & \bh_t = f(\frac{\delta}{|\lambda_w|}{\mbf W} {\mbf h}_{t-1}+ \bU \widetilde{\bX_t}), \label{eq:h2}
\end{eqnarray}
where ${\mbv \mu}$ is a $n_y$-dimensional spatial intercept vector, $\bV_1$ and $\bV_2$ are each $n_y \times n_h$ output weight matrices, and the initial hidden state is set such that $\bh_0 \equiv {\mbf 0}$. Here, we assume that ${\mbf R}_t\equiv \sigma^2_\epsilon \bI$ for all $t$, but note that when necessary, additional temporal or spatial structure can be modeled through the covariance matrix ${\mbf R}_t$ (such additional structure is not needed for the applications presented here). The hidden state parameter, $\lambda_w$, represents the largest eigenvalue of the matrix $\bW$ and $\delta$ is a scaling parameter with a $\text{Unif}(0,1)$ prior. By dividing $\bW$ by $|\lambda_w|$ and restricting $\delta$, we ensure the spectral radius of $\bW$ is at most one. When the spectral radius of $\bW$ exceeds one, the model may exhibit unstable behavior \citep[][]{lukovsevivcius2009reservoir}, and restricting the spectral radius in this fashion is common in the ESN literature, since $\bW$ is not estimated in the ESN model. We find that including $\delta$ in the model provides extra flexibility while providing stability for the hidden states. It is important to note that given the parameters $\delta,\bW$,$\bU$, the initial condition $\bh_0 $, and input vectors, $\widetilde{\bX_t}$ (see below), the hidden states are known and thus, do not need to be directly estimated.

Along with scaling $\bW$, we also extend the traditional RNN model by allowing for additional nonlinearity in (\ref{eq:datastage2}) through $\bh_t^2\equiv(h_{t,1}^2,\dots,h_{t,n_h}^2)'$. By including a nonlinear mapping between the response and $\bh_t$, the proposed model can capture more nonlinear behavior and accommodate more extreme responses \citep[see][]{mcdermott2017ensemble}. It may also be useful to include higher order interactions between the $\bh_t$'s, although such interactions are not helpful for the applications described below. 

We borrow the idea of embedding the input from the dynamical systems literature as introduced by \cite{takens1981detecting}, to define the input vector in (\ref{eq:h2}) as:
\begin{equation}
\widetilde{\bX_t}'=[\bX_t',\bX_{t-\tilde{\tau}}',\dots,\bX_{t-m\tilde{\tau}}']',
\label{embedInput}
\end{equation}
where $\tilde{\tau}$ is usually referred to as the ``embedding lag" and $m$ the ``embedding length," thus leading to a $(m+1)n_x+1$ dimensional input vector (assuming the first element of $\widetilde{\bX_t}'$ corresponds to an intercept term). By embedding the process of interest, the proposed model utilizes all of the recent trajectory of the system, opposed to a single instance in time. Other statistical nonlinear spatio-temporal forecasting methods \citep[e.g.,][]{mcdermott2016model,mcdermott2017ensemble}, have shown that embedding the process of interest can lead to more accurate forecasts and quantifiably better uncertainty measures.

\subsection{BAST-RNN Prior Distributions}
\label{priors}
The presented BAST-RNN model is comprised of multiple high-dimensional parameter weight matrices, resulting in an over-parameterized model. This problem is not unique to the BAST-RNN, and is often a criticism of RNNs and feed-forward neural networks in general. Due to the prevalence of this over-parameterization problem, many solutions have been proposed in the machine learning literature. More recently, a method known as dropout \citep[][]{csrivastava2014dropout,polson2017deep} has shown promise as a tool to deal with over-parameterized weight matrices, thereby helping to prevent over-fitting. In essence, dropout creates a type of ``hard" regularization by removing entire hidden units (and therefore weight parameters) during training. Similarly, ESN models deal with over-parameterized weight matrices by randomly setting a large percentage of parameters in the weight matrices to zero and then drawing the remaining parameters from a bounded or constrained distribution \citep[see][]{mcdermott2017ensemble}. These are just two of the many proposed solutions for regularizing the over-parameterized weight matrices that make up neural network models. For statistical models, addressing this problem is similarly vital to help prevent over-fitting. Therefore, we propose regularization priors for the BAST-RNN model (see below) that borrow ideas from both the dropout and ESN method of regularization.

Allowing for many possible sparse networks is a strength of both the dropout and ESN regularization methods. As discussed throughout the Bayesian machine learning literature \citep[e.g.,][]{mackay1992practical,gan2016scalable}, the natural modeling averaging implicit in the fully Bayesian paradigm acts in a similar way to produce a model averaging effect across many potential networks. Generally in a Bayesian framework, model averaging is induced by using one of the many available priors in the Bayesian variable selection literature \citep[see][for an overview]{o2009review}. For example, stochastic search variable selection (SSVS) priors \citep[][]{george1993variable} represent an effective tool for shrinking parameter values infinitesimally close to zero. In general, SSVS priors consist of a mixture of two distributions, in which one of the distributions shrinks the parameter value near (or to) zero, while the other distribution in the mixture is more vague and allows the parameter to be non-zero.  

Although the traditional SSVS prior uses Gaussian distributions \citep[i.e.,][]{george1997approaches} for the weight matrices $\bW$ and $\bU$, we replace these Gaussian distributions with a truncated normal to create a ``hard" constraint (see (\ref{SSVS1}) and (\ref{SSVS2}) below).  As has been previously noted in the Bayesian neural network literature \citep[i.e.,][]{ghosh2004hierarchical}, the parameters at the top level of the model (i.e.,  $\bV_1\bh_t$ and $\bV_2\bh_t^2$ for the BAST-RNN model in (\ref{eq:datastage2})) are not identifiable. By using truncated normal distributions, we are in some sense constraining the contribution of each weight matrix $\bW$ and $\bU$ towards the $\bh_t$'s. While helping to partly alleviate this identifiability problem, we also find that using truncated normals helps improve mixing when performing MCMC estimation. Finally, as discussed in \cite{ghosh2004hierarchical}, this non-identifiability is not an issue when parameters are given proper priors and interest is only in prediction, as is the case here.

Using the SSVS framework described above, each element of the weight matrix $\bW=\ \{ w_{i,\ell} \}$, for $i=1,\dots,n_h$  and $\ell=1,\dots,n_h$ , is given the following prior distribution: 
\begin{equation}
 w_{i,\ell}=\gamma_{i,\ell}^w \text{TN}_{[-a_w,a_w]}(0,\sigma^2_{w,0})+(1-\gamma_{i,\ell}^w) \text{TN}_{[-a_w,a_w]}(0,\sigma^2_{w,1}),
 \label{SSVS1}
\end{equation}
where $\sigma^2_{w,0}\gg\sigma^2_{w,1}$ and the notation $\text{TN}_{[-a_w,a_w]}$ denotes a truncated normal distribution, truncated between $-a_w$ and $a_w$. Moreover, $\gamma_{i,\ell}^w$ represents an indicator variable with prior, $\gamma_{i,\ell}^w\sim \text{Bernoulli}(\pi_w)$, such that $\pi_w$ can be thought of as the prior probability of including $w_{i,\ell}$ in the model. An analogous prior is used for each element of $\bU=\ \{ u_{i,r} \}$ for $r=1,\dots,(m+1)n_x+1$, such that:
\begin{equation}
u_{i,r}=\gamma_{i,r}^u \text{TN}_{[-a_u,a_u]}(0,\sigma^2_{u,0})+(1-\gamma_{i,r}^u) \text{TN}_{[-a_u,a_u]}(0,\sigma^2_{u,1}), 
 \label{SSVS2}
\end{equation}
where $\gamma_{i,r}^u\sim \text{Bernoulli}(\pi_u)$ and $\sigma^2_{u,0}\gg\sigma^2_{u,1}$. As described in Section \ref{priorChoices}, both hyper-parameters $\pi_w$ and $\pi_u$ are set to small values in order to regularize many of the parameters in the model (since $\sigma^2_{w,1}$ and $\sigma^2_{u,1}$ are set to very small values, as detailed in Appendix A). The priors defined in (\ref{SSVS1}) and (\ref{SSVS2}) mimic aspects of the regularization produced from using dropout or the ESN model by similarly removing or (nearly) zeroing out many of the model parameters. While developing the proposed model we also considered other popular Bayesian variable selection priors such as the Lasso prior \citep[][]{park2008bayesian} and the horseshoe prior \citep[][]{carvalho2010horseshoe}. We found the proposed SSVS priors provided the most flexibility with our approach.

Next, the parameters matrices $\bV_1$ and $\bV_2$ are given traditional SSVS priors with Gaussian distributions (see Appendix A for the full details). Although a $L_2$ (ridge) penalty is typically used for estimating the $\bV$ matrices in the ESN model, we found this penalty to be overly restrictive for the BAST-RNN model. To finish the specification of the model, the spatial intercept is given the Gaussian prior, ${\mbv \mu}\sim \text{Gau}({\mbv 0}, \sigma_\mu^2 \bI)$, and the variance parameter $\sigma^2_\epsilon$ is given the inverse-gamma prior, $\sigma^2_\epsilon \sim \text{IG}(\alpha_\epsilon,\beta_\epsilon)$. See Appendix A for the specific values of the presented hyper-parameters and Section \ref{priorChoices} for a further discussion on certain hyper-parameter choices.

\subsection{Dimension reduction}
\label{dimRed}
With the rise of machine learning and high-dimensional methods has also come the increase in size of spatio-temporal data sets. In most cases, this increase in size can be attributed to the number of spatial locations (or grid-points) in a given data set, rather than the number of time points. When $n_y$ or $n_x$ (or both) is large, the BAST-RNN model can quickly become computationally prohibitive. For example, with more locations, each step of the MCMC algorithm (in particular, Metropolis-Hastings steps) will become more computationally costly. Secondly, with more locations it may be necessary to increase the value of $n_h$, thus increasing the size of all the weight parameter matrices in the model. A common solution to this problem in the spatio-temporal dynamical modeling literature is to use some form of dimension reduction  \citep[i.e.,][]{CandW2011}, which is often justified since the underlying dynamics of such processes typically live in a lower dimensional manifold than the data.

There is a great deal of flexibility when selecting a dimension reduction method for high-dimensional spatio-temporal processes. Depending on the application, any number of methods can be selected from linear methods such as wavelets, splines, or principal components, or nonlinear methods such as Laplacian eigenmaps \citep[][]{laplacianEigen}, restricted Boltzmann machines \citep[][]{nair2010rectified}, or diffusion maps \citep[][]{diffusionMaps}, just to name a few. To describe how dimension reduction can be used with the BAST-RNN model, suppose we let $\bZ_t$ be a $n_z$-dimensional observed response vector at time $t$. Then, for linear dimension reduction, $\bZ_t$ can be decomposed such that $\bZ_t \approx {\mbv \Phi} \bY_t$, where ${\mbv \Phi}$ is a $n_z \times n_b$ basis function matrix and $\bY_t$ is a $n_b$-dimensional vector of basis coefficients. Importantly, we assume that $ n_b\ll n_z$, thus, $\bY_t$ provides a lower-dimensional set of variables (expansion coefficients) with which our model can be built. For example, the proposed BAST-RNN model can be re-formulated using the basis coefficients as follows: 
\begin{eqnarray}
\mbox{data model: }  & \; & {\mbf Z}_t ={\mbv \Phi} \bY_t+ {\mbv \nu}_t \;\; {\mbv \nu}_t \; \sim \; \text{Gau}({\mbf 0},{\mbf \Sigma}_\nu),   \label{eq:datastage3} \\
\mbox{process model: }  & \; & \bY_t={\mbv \mu}+  \bV_1\bh_t + \bV_2\bh_t^2 + {\mbv \epsilon}_t \;\; {\mbv \epsilon}_t \; \sim \; \text{Gau}({\mbf 0},{\mbf R}_t),   \label{eq:processmodel} 
\end{eqnarray}
where the error term ${\mbv \nu}_t$ helps account for the truncated error caused by using a reduced dimension. For some applications it may be more important than others to include the error term (i.e., $ {\mbv \nu}_t$) in (\ref{eq:datastage3}). Although the forecasting applications examined below do not include this truncation error term, the proposed framework allows for the potential to account for such uncertainty.

\section{Computation: Parameter Expansion MCMC}
Similar to non-Bayesian RNN estimation, the nonlinearity and dependence structures in the BAST-RNN model present unique estimation and computational challenges. Both the $\bW$ and $\bU$ weight matrices in the BAST-RNN model are particularly difficult to estimate due to the fact both are within the nonlinear activation function, along with the many dependencies that exist between these two matrices. This dependence occurs since, given the embedded input ($\widetilde{\bX_t}$ above) and $\delta$, the hidden states in (\ref{eq:h2}) are completely determined by $\bW$ and $\bU$. Thus, as $\bW$ and $\bU$ change, so do the values of the hidden states. Importantly, since $\bW$ weights the hidden states, the parameter values of $\bW$ are highly dependent on the specific values of the hidden states and by proxy, the values of $\bU$. 

Parameter expansion data augmentation (PXDA) \citep[e.g.,][]{liu1999parameter,hobert2008theoretical,hobert2011data} is a method developed for missing data problems in which mixing for MCMC algorithms is difficult due to dependencies between parameters. While the parameter expansion in the PXDA algorithm is generally applied to missing data, we borrow the parameter expansion idea and apply it to the sampling of the $\bW$ matrix (we did not find it necessary to use this same technique on the $\bU$ matrix), since $\bW$ directly weights the hidden states. In essence, parameter expansion MCMC (PX-MCMC) introduces extra parameters (referred to as the expansion parameters) into the model to create extra randomness. For example, suppose for a given iteration of the MCMC algorithm we sampled $\bW$ and then $\bU$. Instead of moving from $\bW$ to $\bU$ (i.e., $\bW \rightarrow \bU$), the expanded parameter is used to create an intermediate step such that $\bW  \rightarrow \bW^*  \rightarrow \bU$. That is, $\bW^*$ is a randomly transformed version of $\bW$, thus helping to break some of the dependence between weight matrices $\bW$ and $\bU$ \citep[][refers to this randomness as a ``shake-up" of the parameters]{hobert2011data}.  Without this extra randomness, samples for the weight matrices  $\bW$ and $\bU$ quickly become degenerate.  

By introducing this intermediate step, the mixing in the MCMC algorithm greatly improves for both the $\bW$ and $\bU$ matrix. The amount of randomness used to transform $\bW$ into $\bW^*$ can be thought of similarly to the learning rate parameter used in traditional stochastic gradient descent (SGD) or SG-MCMC algorithms for machine-learning problems. Typically, a learning rate parameter is used in SGD algorithms to determine how fast or slow the weights in a given model are learned. For example, in SG-MCMC, at each iteration when the model parameters are updated, a standard multivariate Gaussian distribution multiplied by a learning rate parameter is added to the updated parameter values \citep[i.e.,][]{gan2016scalable}. Analogous to the learning rate for SGD, the extra randomness induced by $\bW^*$ allows the algorithm to better search the entire parameter space, thus improving the mixing of the algorithm. 

To more rigorously describe the PX-MCMC algorithm, we need to define additional notation. Suppose we introduce the expansion parameter matrix ${\mbv \alpha }$,  where ${\mbv \alpha}=\{\alpha_{i,\ell}\}$  for $i=1,\dots,n_h$ and $\ell=1,\dots,n_h$, and ${\mbv \alpha} \subset \bA$, where $\bA \in \R^{n_h^2}$. Next, we define the transformation $t_{\mbv \alpha} : \bW \longrightarrow \bW$, where we require $t_{\mbv \alpha}$ to be a one-to-one differentiable function and denote the Jacobian for this transformation as $J_{\mbv \alpha}(\bW)$. Let ${\mbv\Gamma}_{V_1}, {\mbv \Gamma}_{V_2}, {\mbv \Gamma}_U,$ and ${\mbv \Gamma}_W$ denote all of the indicator variables for the SSVS priors corresponding to the respective weight matrices in (\ref{eq:datastage2}) and (\ref{eq:h2}). We define $\Theta$ to be all of the parameters in the model not associated with $\bW$; that is, $\Theta\equiv \{ {\mbv \mu},\bV_1,\bV_2, {\mbv\Gamma}_{V_1}, {\mbv \Gamma}_{V_2}, \bU, {\mbv \Gamma}_U, \delta, \sigma^2_\epsilon  \}$. Furthermore, let $\bY_{1:T}\equiv \{\bY_1,\dots,\bY_T\}$ and $\widetilde{\bX}_{1:T}\equiv \{ \widetilde{\bX}_1,\dots,\widetilde{\bX}_T\}$. Finally, we define the likelihood of the model (before the introduction of the expansion parameter matrix ${\mbv \alpha }$) using the following slight abuse of notation $\prod\limits^T_{t=1} [\bY_t \mid \Theta, \bW,{\mbv \Gamma}_W,\widetilde{\bX}_t ] = [\bY_{1:T} \mid \Theta, \bW,{\mbv \Gamma}_W,\widetilde{\bX}_{1:T} ]$, where $[\cdot]$ denotes a distribution.

Now, we outline the PX-MCMC for the BAST-RNN model; note, we leave the detailed description of the presented algorithm for Appendix B. Using the notation defined above, one can show (see Appendix B) the following relationship for the joint posterior of $\bW$ and ${\mbv \Gamma}_W$:
\begin{equation}
[\bW,{\mbv \Gamma}_W \mid \Theta,\bY_{1:T},\widetilde{\bX}_{1:T}]= \int_\bA  [t_{\mbv \alpha}(\bW),{\mbv \Gamma}_W\mid \Theta,\bY_{1:T},\widetilde{\bX}_{1:T}] \ | J_{\mbv \alpha}(\bW)| \ [{\mbv \alpha}] \ d{\mbv \alpha}.
 \label{posterior1}
\end{equation}
To sample from the integral in (\ref{posterior1}), we assume $\bW', {\mbv \Gamma}_W \sim [t_{\mbv \alpha}(\bW),{\mbv \Gamma}_W\mid \Theta,\bY_{1:T},\widetilde{\bX}_{1:T}]$. We will take $\bW=t_{{\mbv \alpha}}^{-1}(\bW')$, thus allowing for the joint sampling of $\bW$ and ${\mbv \Gamma}_W$, leading to step 1 in Algorithm 1. Next, the draw from $[{\mbv \alpha}]$ is denoted as ${\mbv \alpha}_0$, thus step 2 in Algorithm 1.

We assume ${\mbv \alpha} \sim \text{Gau}({\mbv 0}, \sigma^2_\alpha \bI)$ for the BAST-RNN model implementation, where the prior variance $\sigma^2_\alpha$ can be thought of analogous to the learning rate parameter used in many machine learning estimation algorithms. There is a great deal of flexibility with regards to the particular distribution used for $[{\mbv \alpha}]$ \citep[i.e.,][]{hobert2011data} and its choice should depend on the particular model and application. Letting $\widetilde{\bW}\equiv t_{{\mbv \alpha}_0}^{-1}(\bW)$, we can sample ${\mbv \alpha}$ and $\Theta$ using the following full-conditional distributions (see Appendix B for further details) :

\begin{eqnarray}
 & \; &     [ {\mbv \alpha} \mid  t_{\mbv \alpha} (\widetilde{\bW }),{\mbv \Gamma}_W,\Theta, \bY_{1:T},, \widetilde{\bX}_{1:T}  ]   \propto \nonumber \\ 
& \; &   [\bY_{1:T} \mid  t_{\mbv \alpha} (\widetilde{\bW }),{\mbv \Gamma}_W, \Theta,{\mbv \alpha},\widetilde{\bX}_{1:T}  ]   [  t_{\mbv \alpha} (\widetilde{\bW }) \mid {\mbv \Gamma}_W, {\mbv \alpha}   ] \ [{\mbv \alpha} ]  \ |  J_{\mbv \alpha} (\widetilde{\bW }) | ,  
 \label{alphaFC}
     \end{eqnarray}
               
                              \vspace{-1.75cm} 

                \begin{align}
                [ \Theta \mid  t_{\mbv \alpha} (\widetilde{\bW }),{\mbv \Gamma}_W,{\mbv \alpha}, \bY_{1:T}, \widetilde{\bX}_{1:T}  ] 
               & \propto   [\bY_{1:T} \mid  t_{\mbv \alpha} (\widetilde{\bW }), \Theta,\widetilde{\bX}_{1:T}  ] \ [\Theta]. 
                \label{thetaFC}
                \end{align}

Taking the previous three equations together, we can form the PX-MCMC algorithm given in Algorithm 1. For the sake of brevity, we leave the specific full-conditional distributions for all the model parameters for Appendix C.

\begin{algorithm}[H]
\singlespace
\begin{enumerate}
\item Sample $\bW,{\mbv \Gamma}_W$ from: $ [\bW,{\mbv \Gamma}_W\mid \Theta,\bY_{1:T},\widetilde{\bX}_{1:T} ] \propto$ 

\vspace{1mm} 

$[\bY_{1:T}\mid  \Theta, \bW,{\mbv \Gamma}_W, \widetilde{\bX}_{1:T}  ] \ [\bW\mid {\mbv \Gamma}_W] \ [{\mbv \Gamma}_W]$.
\item Generate $\alpha_{0,i,\ell}\sim \text{Gau}(0, \sigma^2_\alpha )$ for $i=1,\dots,n_h$ and $\ell=1,\dots,n_h$. 
\item Transform $\widetilde{\bW}=t^{-1}_{{\mbv \alpha}_0}(\bW)$.
\item Sample ${\mbv \alpha}$ from: $[ {\mbv \alpha} \mid  t_{\mbv \alpha} (\widetilde{\bW }),{\mbv \Gamma}_W,\Theta, \bY_{1:T} ,\widetilde{\bX}_{1:T}] \propto $

\vspace{1mm}

$[\bY_{1:T} \mid  t_{\mbv \alpha} (\widetilde{\bW }),{\mbv \Gamma}_W, \Theta,{\mbv \alpha},\widetilde{\bX}_{1:T}  ] \  [  t_{\mbv \alpha} (\widetilde{\bW }) \mid {\mbv \Gamma}_W, {\mbv \alpha}   ] \ [{\mbv \alpha} ]  \ |  J_{\mbv \alpha} (\widetilde{\bW }) | $.
\item Sample $\Theta$ from: $ [ \Theta \mid  t_{\mbv \alpha} (\widetilde{\bW }),{\mbv \Gamma}_W,{\mbv \alpha}, \bY_{1:T}, \widetilde{\bX}_{1:T}  ] \propto   [\bY_{1:T} \mid  t_{\mbv \alpha} (\widetilde{\bW }), \Theta,\widetilde{\bX}_{1:T}  ] \ [\Theta].$
\end{enumerate}

\vspace{3mm}

\caption{PX-MCMC algorithm}

\end{algorithm}


\section{Applications}
We begin by discussing the specifics of model implementation, including comparison metrics and methods, the MCMC setup, and specific hyper-parameter choices. We then present the analysis of a simulated multiscale Lorenz data set from the Lorenz dynamical system. In addition, the setup and results of a Pacific SST long-lead forecasting problem are given, followed by an application to state-level unemployment data in the U.S..

\subsection{Validation Measures and Alternative Models}
\label{modelsetup}
Since the stated goal of developing the BAST-RNN model is to produce accurate forecasts with realistic uncertainty bounds, we evaluate the model in terms of both mean squared prediction error (MSPE) and continuous ranked probability score (CRPS). Both measures are only calculated for out-of-sample values, since the focus of the model is on forecasting. For our purposes, the MSPE is defined as the average squared difference between the out-of-sample forecasts and true out-of-sample values across all time periods and spatial locations. Moreover, for a predictive CDF $F$ and true out-of-sample realization $h$, CRPS is defined as \cite[e.g.,][]{matheson1976scoring}:
\begin{equation}
\text{CRPS}(F,h)= \int_ \R (F(r)- \mathbbm{1}\{ r \geq h \} )^2 dr.
\end{equation}
The usefulness of CRPS lies in its ability to both quantify the accuracy and distribution of a forecast, thus producing a principled (proper scoring rule) measure of how well a model quantifies uncertainty \citep[i.e.,][]{gneiting2014probabilistic}. In all the applications presented below, after the model is trained on in-sample data, out-of-sample forecasts are generated successively at the given lead time. We define lead time as the temporal difference between the input and the response. These successive forecasts are made by repeatedly plugging in the inputs for a given lead time to get out-of-sample forecasts, using the posterior samples from the BAST-RNN.

For the sake of comparison, we also evaluated the ensemble quadratic ESN (E-QESN) model from \cite{mcdermott2017ensemble} for all of the applications below. The E-QESN model presents a strong comparison model since it can also quantify uncertainty and shares much of the same flexibility as the BAST-RNN model. Few other methods share the E-QESN's ability to produce forecasts with uncertainty quantification at such a low computational and implementation cost.

We also compared to a model referred to as the linear DSTM \citep[e.g.,][Chapter 7]{CandW2011}, defined here as:
\begin{equation}
Y_{t,i} = \sum_{j=1}^{n_y} a_{ij} Y_{t-1,j} + \zeta^{(l)}_{t,i},
\label{linDSTM}
\end{equation}
for each location $Y_{t,i}$, where $\{ a_{i,j}\}$ are weight parameters and $\zeta_{t,i}^{(l)}$ is a spatially referenced noise term, such that ${\mbv \zeta}^{(l)}_{t}\sim\text{Gau}({\mbv 0},{\mbv \Sigma}_{\zeta_l})$. Finally, we compare to the GQN model discussed above \citep[][]{wikle2010general}. For the results presented below, the GQN model is defined as:
\begin{equation}
Y_{t,i} = \sum_{j=1}^{n_y} a_{ij} Y_{t-1,j} + \sum_{k=1}^{n_y} \sum_{\ell=1}^{n_y} b_{i,k,\ell} Y_{t-1,k}Y_{t-1,\ell}+ \zeta^{(q)}_{t,i}, 
\label{GQN}
\end{equation} 
where ${\mbv \zeta}^{(q)}_{t}\sim\text{Gau}({\mbv 0},{\mbv \Sigma}_{\zeta_q})$. Both ${\mbv \Sigma}_{\zeta_l}$ and ${\mbv \Sigma}_{\zeta_q}$ are estimated empirically using the residuals from the training period. Although both the GQN and linear DSTM can be formulated as Bayesian models, such formulations are not pursued here. Instead, forecast distributions are calculated through a Monte Carlo approach for both models. While this is not an exhaustive list of comparison methods, these methods represent much of the state-of-the art in statistical spatial-temporal modeling and nonlinear spatial-temporal forecasting.
 
\subsection{BAST-RNN Implementation Details}
\label{priorChoices}

Note, the implementation settings discussed here are used for all of the presented applications, with slight deviations for specific applications as discussed below. The BAST-RNN model is implemented using the PX-MCMC algorithm (Algorithm 1), sampling 100,000 iterations with the first 25,000 iterations treated as burn-in, while thinning the samples such that every fifth post burin-in sample was retained. We monitored convergence by examining the trace plots for the parameters in the model along with the posterior forecasts (a sample of such trace plots can be found in Appendix D). The number of hidden units ($n_h$) is set to 20. We found this number of hidden units balanced computational cost and forecast accuracy in that larger numbers of hidden units produced similar results in terms of forecast accuracy, but substantially slowed the algorithm.  Although not pursued here, the number of hidden units could be varied by using advanced computational methods such as reversible jump MCMC \citep[][]{green1995reversible}. Selection of the parameters for the embedded input, defined in (\ref{embedInput}), is conducted by using cross-validation, over an application specific grid with the E-QESN model \citep[see][for a detailed description of this procedure]{mcdermott2017ensemble}. As suggested by \cite{lukovsevivcius2012practical}, both the input and response are scaled by their respective means and standard deviations.

 While we leave the specific hyper-parameter values used in the prior distributions to Appendix A, we will briefly discuss the more important of these choices. Specifically for the parameter weight matrices that make up the hidden units (i.e., $\bW$ and $\bU$), the hyper-parameters $\pi_w$ and $\pi_u$ (as defined in Section \ref{priors}) are set to small values to encourage sparseness and prevent overfitting. In particular, these hyper-parameters are set such that $\pi_w> \pi_u$, since the matrix $\bU$ is weighting the data; we found this specification helped prevent overfitting to the in-sample data. Moreover, $a_w$ and $a_u$ are both set to small values so that $a_w=a_u$, as to follow the common practice in machine learning of bounding parameter values to prevent overfitting \citep[i.e.,][]{lukovsevivcius2012practical}. 

\subsection{Simulation: Multiscale Lorenz-96 Model}
Many RNN methods in the literature use the classic three-variable Lorenz model from \cite{lorenz1963deterministic} to evaluate forecasting ability \citep[e.g.,][]{ma2007chaotic,chandra2012cooperative}. Due to the chaotic and nonlinear behavior of the Lorenz model, this system produces data resembling a realistic nonlinear forecasting problem, but it has an unrealistically low state dimension (three) and is not spatially referenced. Here, evaluation of the BAST-RNN model is applied to a less cited, but more spatially interesting Lorenz model \citep[][]{lorenz1996predictability}, often referred to as the Lorenz-96 model, which explicitly includes spatial locations and structure. In particular, we consider    a more complicated extension of the Lorenz-96 model, the {\it multiscale Lorenz-96} model, that contains interacting large-scale and small-scale processes, where the large-scale locations are directly influenced by neighboring small-scale locations and vice-versa \citep[e.g.,][]{wilks2005effects,chorin2015discrete,grooms2015framework}. 

While multiple parameterizations exist for the multiscale Lorenz-96 model, we use the following parameterization from \cite{chorin2015discrete} (note, the superscript $L$ is used throughout to signify variables from the  Lorenz-96 model):
\begin{eqnarray}
& \; & \frac{dx^L_{k_L}}{dt} = x^L_{k_L-1}(x^L_{k_L+1} - x^L_{k_L-2}) - x^L_{k_L} + \tilde{F}+ \frac{h_x}{J_L} \sum_{j_L} y^L_{j_L,k_L}+\eta^{(1)}_{k_L},   \label{eq:L96} \nonumber \\
& \; &   \frac{y^L_{j_L,k_L}}{dt} =\frac{1}{\epsilon_L}[ y^L_{j_L+1,k_L}(y^L_{j_L-1,k_L}-y^L_{j_L+2,k_L})-y^L_{j_L,k_L}  + h_y x^L_{k_L} ],
\end{eqnarray}
for $j_L=1,\dots,J_L$ and $k_L=1\dots,K_L$ (for notational convenience the subscript $t$ has been suppressed from (\ref{eq:L96})). The state variable $x_{k_L}^L$ denotes the process at a large-scale process location, with each large-scale location having $J_L$ corresponding small-scale locations denoted by the process $y_{j_L,k_L}^L$. Each of the large-scale locations can be thought of as equally spaced spatial variables on a one-dimensional circular spatial domain such that $x^L_{K_L+1}=x^L_{1}$ (i.e., periodic boundary conditions). A given set of small-scale locations corresponding to a particular large-scale location is defined with a similar spatial domain and boundary condition. 

The parameter $\tilde{F}$ in (\ref{eq:L96}) denotes a forcing parameter, while $\epsilon_L$ controls the time-scale separation between the large and small-scale processes, $\eta^{(1)}_{k_L}$ is an additive independent Gaussian noise term such that $\eta^{(1)}_{k_L}\overset{iid}{\sim}\text{Gau}(0,\sigma^2_{\eta_1})$, with $\sigma^2_{\eta_1}=1$, and $h_x$,  $h_y$ control how much the large and small-scale locations influence each other, respectively. For the analysis using the BAST-RNN model, we simulate from the full model but treat the small-scale locations as unobserved and evaluate the BAST-RNN only on the large-scale locations, thus creating a difficult but realistic nonlinear spatio-temporal forecasting problem.  After burn-in, 400 time periods are retained from the the multiscale Lorenz-96 model, with the last 75 time periods treated as out-of-sample. The data are simulated with a time step of $\Delta=.05$ using an Euler solver. We use the same parameter values as \cite{chorin2015discrete} to simulate the data: $K_L=18,J_L=20,\tilde{F}=10,\epsilon_L=0.5,h_x=-1,$ and $h_y=1$. In order to create a more statistically-oriented forecasting problem, Gaussian white noise error is added to each large-scale realization, so that $z^L_{k_L}=x^L_{k_L}+\eta^{(2)}_{k_L}$, where $\eta_{k_L}^{(2)}\overset{iid}{\sim} \text{Gau}(0,\sigma^2_{\eta_2})$, with $\sigma_{\eta_2}^2=(2.5)^2$. In addition, the forecasting problem is made slightly more nonlinear by setting the lead time to three periods (i.e., the input and response are separated by three periods). Along with using the implementation settings detailed in Section \ref{priorChoices}, the embedded input parameters $\tilde{\tau}=2$ and $m=4$ are used.

\begin{figure}[H]
  \centering
    \captionsetup{font=footnotesize}
\includegraphics[width=16cm,height=10cm]{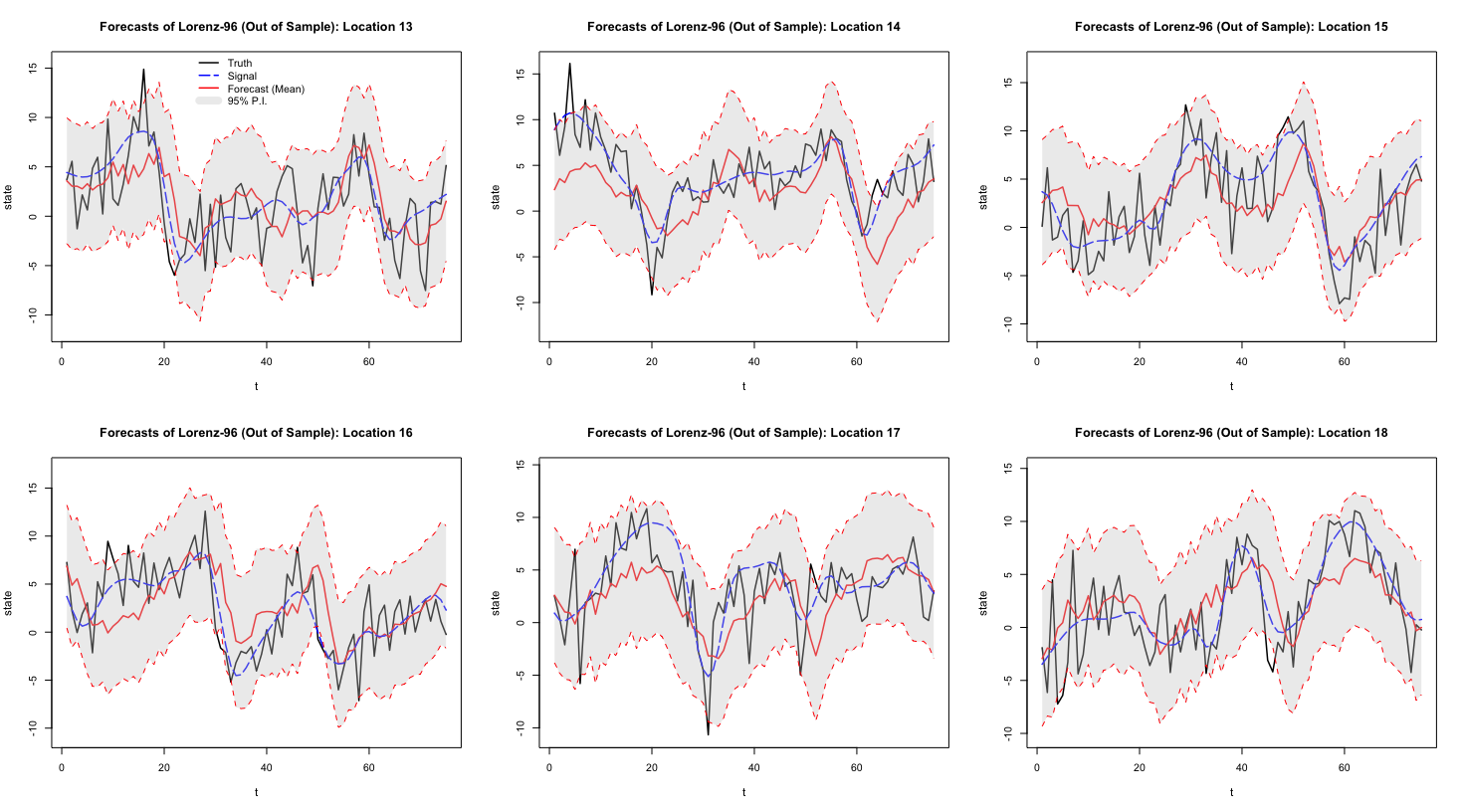}
\caption{Posterior out-of-sample summaries for 6 of the 18 large-scale locations from the simulated multiscale Lorenz-96 data over 75 periods. The black line in each plot represents the true simulated value, while the red line denotes the forecasted posterior mean from the BAST-RNN model. The blue dashed line denotes the true signal of the process, defined to be the value of the large-scale locations in equation (\ref{eq:L96}) before the additive error, $\text{Gau}(0,\sigma^2_{\eta_1})$, is applied. The shaded grey area in each plot signifies the 95\% posterior prediction intervals.}  
\label{fig:Figure_1}
\end{figure}

Posterior mean forecasts and prediction intervals (P.I.s) for the BAST-RNN model with the multiscale Lorenz-96 data are shown for six locations in Figure \ref{fig:Figure_1}. Note that because a low signal-to-noise ratio was used to simulate the data, the true signal is substantially corrupted by the additive noise (as shown by the blue dotted line used to represent the true signal of the process in Figure \ref{fig:Figure_1}). Despite the high level of noise, the model recovers much of the signal for the six locations shown in Figure \ref{fig:Figure_1}. Moreover, it appears that many of the true values of the process are captured by the 95\% P.I.s. Across all 18 large-scale locations in the simulated data, 95.1\% of the true values are contained within the 95\% P.I.s, while only 86.4\% of the true values are contained within the intervals produced by the E-QESN model.

A more detailed comparison of the BAST-RNN model and the three models described in Section \ref{modelsetup} can be found in Table 1. In particular, Table 1 shows the BAST-RNN outperforming the other three competing models by producing lower values for the MSPE. It is not entirely surprising that the BAST-RNN and E-QESN model outperformed the other two less flexible models considering the level of nonlinearity in the simulated data. In addition, compared to the E-QESN model, Table 1 also shows the BAST-RNN model produces superior uncertainty measures based on a lower CRPS. Overall, these results simultaneously demonstrate the ability of the BAST-RNN model to accurately forecast the trajectory of the states in a nonlinear process, while also giving robust measures of uncertainty.

\begin{table}[H]
\begin{varwidth}{2.15in}
\begin{framed}
    \captionsetup{font=scriptsize }
\label{Table_1}
\footnotesize
\begin{tabular}{ccccccccc} 
Model &  MSPE  &  CRPS  \\
\hline
BAST-RNN & 13.09 &  154.46 \\ 
E-QESN & 13.99 & 166.34  \\ 
GQN & 14.85  & 172.50   \\
Lin. DSTM & 15.11 & 166.60    \\
\end{tabular}
\caption{Comparison for the 4 forecasting methods for the simulated multiscale Lorenz-96 data in terms of mean squared prediction error (MSPE) and continuous ranked probability score (CRPS), with a lead time of three periods. Both metrics are calculated over all out-sample periods and locations.}
\end{framed}
\end{varwidth}
\label{tab:Table_1}
\end{table}

\subsection{Application: Long-lead Tropical Pacific SST Forecasting}
Tropical Pacific SST is one of the largest sources of variability affecting global climate \citep[e.g., see the overview in][]{hu2017extreme}. Changes in SST at various time-scales contribute to extreme weather events across the globe, from hurricanes to severe droughts, as well as related impacts (e.g., waterfowl migration). Therefore, accurate long-lead SST forecasts are vital for many resource managers. Of particular interest when considering SST is the anomalous warming (El Ni$\tilde{\text{n}}$o) and cooling (La Ni$\tilde{\text{n}}$a) of the Pacific ocean, referred to collectively as the El Ni$\tilde{\text{n}}$o Southern Oscillation (ENSO) phenomena.

The focus of our analysis is on the ENSO phenomena that occurred during 2015 and 2016. Besides being one of the most extreme ENSO events on record, many forecasting methods that were effective for past ENSO cycles failed to accurately forecast the 2015-2016 cycle \citep[i.e.,][]{l2016observing,hu2017extreme}. As described in \cite{barnston2012skill}, there are currently a suite of both deterministic and statistical methods for forecasting SST, with the statistical models often performing as well or better than the deterministic models. A summary of the deterministic models used to forecast SST can be found in works such as \cite{barnston1999predictive} and \cite{jan2005did}. Some nonlinear statistical models that have shown success for the ENSO forecasting problem include a general quadratic nonlinear (GQN) model \citep[i.e.,][]{wikle2010general}, a switching Markov model \citep[][]{berliner2000long}, and classic neural network models \citep[i.e.,][]{tangang1998forecasting}; for a more expansive list of nonlinear SST models see \cite{mcdermott2017ensemble}. It is important to note that to our knowledge, this is the first RNN method applied to the ENSO forecasting problem with a formal mechanism for quantifying uncertainty. Furthermore, the BAST-RNN model is used to produce out-of-sample forecasts and P.I.s with a lead time of six months for the 2015-2016 ENSO cycle. Due to the lead time and intensity of the 2015-2016 ENSO cycle, this presents a difficult nonlinear forecasting problem.

\begin{figure}[H]
  \centering
    \captionsetup{font=footnotesize}
\includegraphics[width=10cm,height=7.5cm]{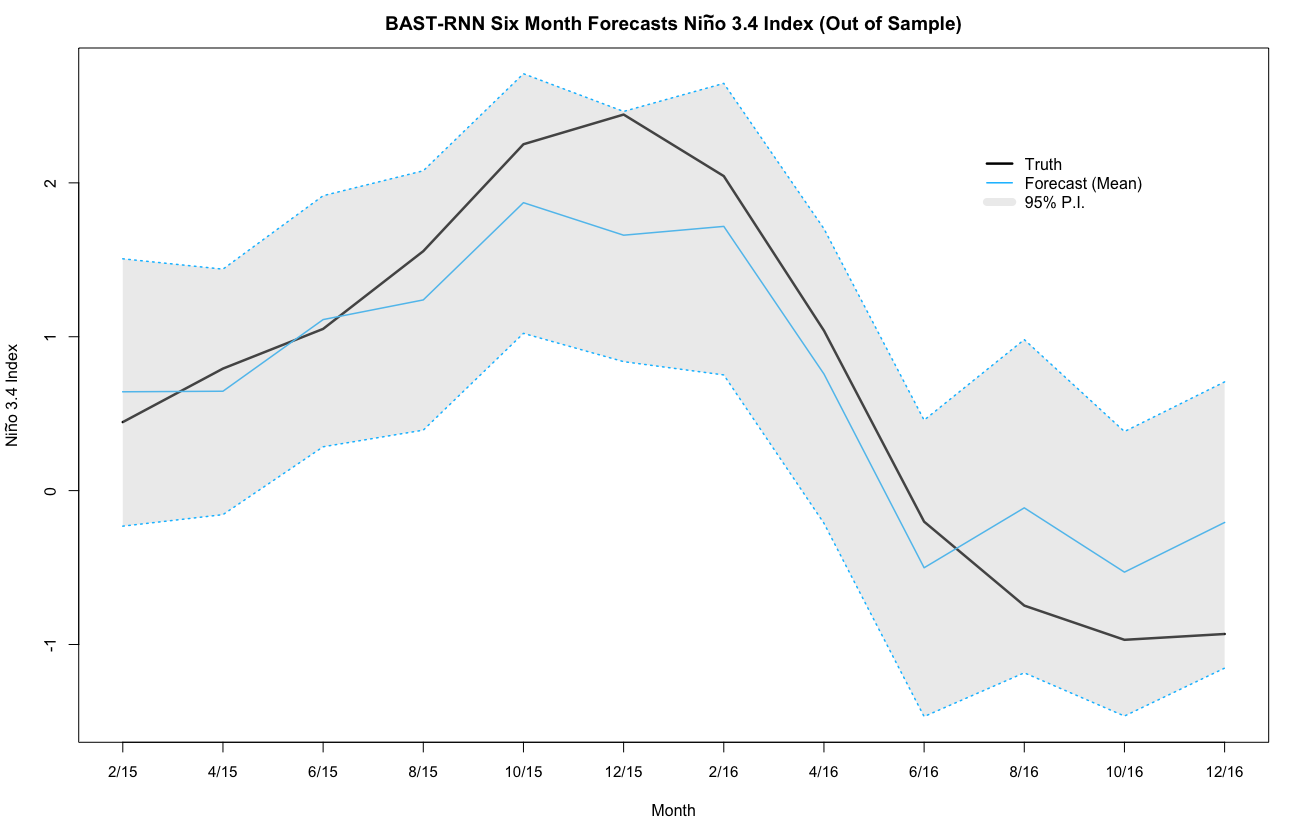}
\caption{Summary of the posterior results with the BAST-RNN model for the Ni\~no 3.4 index. For a given month, the Ni\~no 3.4 index is defined as the average SST over all locations in the Ni\~no 3.4 region ($5^{\circ}$S-$5^{\circ}$N,$120^{\circ}$-$70^{\circ}$W). The solid black lines denotes the true Ni\~no 3.4 index for a given month during the 2015-2016 cycle. Posterior mean out-of-sample forecasts from the BAST-RNN model are denoted by the light blue line. The grey shaded area represents the 95\% P.I.s from the BAST-RNN for the Ni\~no 3.4 index. All values are given in units of degree Celsius.}  
\label{fig:Figure_3}
\end{figure}

The SST forecasting application uses monthly data over a spatial domain covering $29^{\circ}$S-$29^{\circ}$N latitude and $124^{\circ}$E-$70^{\circ}$W longitude, with a resolution of $2^{\circ} \times 2^{\circ} $, leading to 2,229 oceanic spatial locations. The data set is available from the publicly available extended reconstruction sea surface temperature (ERSST) data (\url{http://iridl.ldeo.columbia.edu/SOURCES/.NOAA/}) provided by National Ocean and Atmospheric Administration (NOAA) and covers a period from 1970 through 2016. As is common in the climatology literature, the SST data are converted into anomalies by subtracting the monthly climatological means calculated (in this case) over the period 1981--2010, for each spatial location. Furthermore, when constructing ENSO forecasting methods, it is common to evaluate the performance of a given method using the univariate summary measure for ENSO referred to as the Ni\~no 3.4 index. Much of the variability in the ENSO phenomena is contained in the Ni\~no 3.4 region (i.e., $5^{\circ}$S-$5^{\circ}$N,$120^{\circ}$-$70^{\circ}$W), so for our purposes, the Ni\~no 3.4 index is simply the average SST anomaly over all locations in this region for a given month (see Figure \ref{fig:Figure_3}).

Training of the model is implemented using Algorithm 1 and the setup from Section \ref{priorChoices} with data from January 1970-August 2014, while out-of-sample forecasts were made every two months for a period from February 2015-December 2016 (i.e., the 2015-2016 ENSO cycle) with a lead time of six months. Using the description in Section \ref{dimRed}, dimension reduction is conducted using empirical orthogonal functions (EOFs), also referred to as spatial-temporal principal components \citep[see Chapter 5 of][]{CandW2011}, on both the input and response. The first 10 EOFs, which account for over 80\% of the variability in the data, are retained for both the input and response. This same number of EOFs has been used in multiple previous SST studies \citep[i.e.,][]{berliner2000long,gladish2014physically}. Note, the first two EOFs account for almost $57\%$ of the variation in the data. Due to this, some of the variables associated with these EOFs were given higher values for $\pi_u$ (see Appendix A for the specific values and a more detailed discussion of these choices). Once again, the embedded input parameters are selected using the E-QESN model such that $\tilde{\tau}=6$ and $m=4$.

\begin{figure}[H]
  \centering
    \captionsetup{font=footnotesize}
\includegraphics[width=10cm,height=9cm]{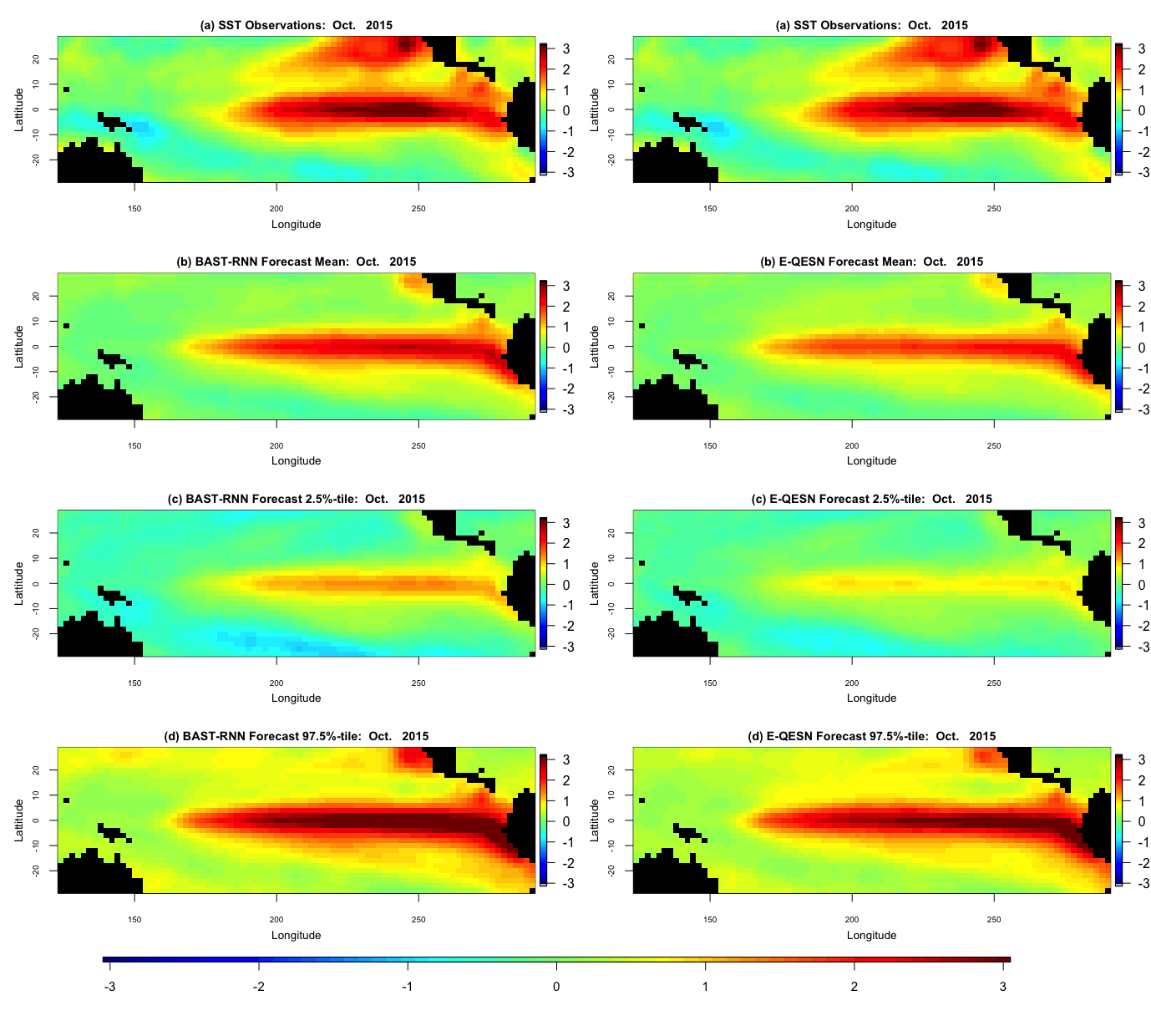}
\caption{Spatial posterior summaries of SST anomalies for all 2,229 oceanic spatial locations in the SST long-lead forecasting application for October 2015. The left column shows results from the BAST-RNN model, while the right column contains results from the competing E-QESN model for the same period. (a) The true SST for October 2015. (b) Posterior mean out-of-sample forecasts for the BAST-RNN model and mean out-of-sample forecasts over all ensembles for the E-QESN model. (c) Lower 2.5\%  point wise P.I.s for the respective forecasting method. (d) Upper 97.5\% point wise P.I.s for the respective forecasting method. All plots are in units of degree Celsius. }  
\label{fig:Figure_2}
\end{figure}

Comparison of the forecasting ability of the BAST-RNN model and the E-QESN model for the entire spatial domain can be seen in Figure \ref{fig:Figure_1} for October 2015. Occurring directly before the peak of the 2015-2016 ENSO cycle (see Figure \ref{fig:Figure_3}), October 2015 represents an important month from the most recent ENSO cycle.  Overall, both methods capture much of the warm intensity trend, with the BAST-RNN model forecasting a slightly higher intensity (especially for the Ni\~no 3.4 region) compared to the E-QESN method. Although both methods appear to produce P.I.s with similar widths, the BAST-RNN model correctly indicates the possibility of a more intense warm event than the E-QESN model. Importantly, the highest intensity true values from the Ni\~no 3.4 region for October 2015  are contained within the 95\% P.I.s for the BAST-RNN model. 

Furthermore, the BAST-RNN model is evaluated in terms of the previously described Ni\~no 3.4 index in Figure \ref{fig:Figure_3}. Much of the overall temporal nonlinear trend of the 2015-2016 ENSO cycle is captured by the BAST-RNN model, as shown in Figure \ref{fig:Figure_3}, with all of the true values contained within the 95\% P.I.s. We should note, like the vast majority of ENSO forecasting models, the forecast mean from the BAST-RNN model also underestimates the peak of the ENSO cycle. Considering the 2015-2016 ENSO cycle was one of the most extreme ENSO cycles of recent record \citep[i.e.,][]{l2016observing}, it is not entirely surprising that most models underestimated the peak of the cycle. But, it is important to reiterate that the true peak was contained in the out-of-sample forecast P.I.s for the BAST-RNN model, unlike many other ENSO forecast models.

Once again we evaluate the performance of the BAST-RNN against the three comparison models described above. Throughout the 2015-2016 ENSO cycle, Table 2 shows the BAST-RNN as a more accurate long-lead forecasting model than the other three models. The BAST-RNN model greatly outperforms the other models over the Ni\~no 3.4 region, illustrating the model's ability to forecast nonlinear dynamics. Moreover, Table 2 also shows the BAST-RNN model has the lowest CRPS over all 2,229 locations in the application, thus providing useful uncertainty information across the entire spatial domain. By producing sensible uncertainty metrics for events six months into the future, the BAST-RNN model gives resource managers advanced information on which informed decisions can be made. Considering the widespread impact SST has on the global climate, such reliable information is invaluable from both a scientific and economical perspective.

\begin{table}[H]
\begin{varwidth}{4.75in}
\begin{framed}
    \captionsetup{font=scriptsize }
\footnotesize
\begin{tabular}{ccccccccc} 
Model & Overall MSPE  & Ni\~no 3.4 MSPE &  CRPS &   Ni\~no 3.4 CRPS  \\
\hline
BAST-RNN & 0.248 &  0.193 &  3.404  & 0.290 \\ 
E-QESN &   0.288  &  0.261 & 3.722 & 0.354\\ 
GQN & 0.309 & 0.619 & 3.924  & 0.538  \\
Lin. DSTM & 0.328 & 0.785 & 3.752  & 0.699  
\label{Table_2}
\end{tabular}
\caption{Summary metrics for each of the four methods evaluated for the long-lead SST forecasting application. Overall MSPE denotes the MSPE calculated over all out-of-sample periods and all oceanic locations. The column labeled CRPS denotes the CRPS calculated over all locations and out-of-sample time periods.The columns Ni\~no 3.4 MSPE and Ni\~no 3.4 CRPS denote the MSPE and CRPS, respectively, over all locations in the Ni\~no 3.4 region and all out-of-sample periods. }
\end{framed}
\end{varwidth}
\end{table}

\subsection{Application: U.S. State-Level Unemployment Rate}
Finally, the BAST-RNN model was applied to forecasting of state unemployment rates in the Midwest of the U.S.. Previous research by  \cite{liang2005bayesian} and \cite{sharma2016unemployment} have shown neural network models to be successful for forecasting national unemployment. A lesser studied, but equally important, component of unemployment forecasting is the spatio-temporal problem of forecasting state-level rates. Moreover, while linear models have shown success at forecasting unemployment rates at short lead times, nonlinear models generally produce more accurate results at longer lead times \citep[i.e.,][]{terasvirta2005linear,liang2005bayesian}. The BAST-RNN can account for the nonlinearity present for a longer lead forecast, while also incorporating the dependence between unemployment rates in near-by states. 

Compared to the previous two applications (Lorenz system and SST), the U.S. unemployment rate is a much slower moving process (i.e., compare Figure \ref{fig:Figure_3} and \ref{fig:Figure_4} where each displays approximately one (quasi) cycle of the processes of interest, and note that Figure \ref{fig:Figure_4} covers a temporal span twice as long as Figure \ref{fig:Figure_3}). For example, there have been many fewer U.S. unemployment cycles over the past 40 years compared to ENSO cycles. Due to this difference in the rate of the dynamical process, smaller values for the hyper-parameters $a_w$ and $a_u$ are used for the unemployment application (see Appendix A for the specific values and a more detailed discussion). By using smaller values for $a_w$ and $a_u$ the model has more memory of recent past values \citep[i.e.,][]{lukovsevivcius2009reservoir}, which is necessary for slower moving processes. Similar to the two previous applications, the embedding parameters were selected with the E-QESN model, with the model selecting $\tilde{\tau}=0$  and $m=0$ (i.e., $\widetilde{\bX_t}'=\bX_t$ in (\ref{embedInput})).

\begin{table}[H]
\begin{varwidth}{2.15in}
\begin{framed}
    \captionsetup{font=scriptsize }
\footnotesize
\begin{tabular}{ccccccccc} 
Model &  MSPE  &  CRPS  \\
\hline
BAST-RNN & 0.605 &  26.89 \\ 
E-QESN & 0.902 & 35.34  \\ 
GQN & 0.964  &  37.29   \\
Lin. DSTM & 0.867 & 33.66   
\label{Table_3}
\end{tabular}
\caption{Comparison for the 4 forecasting methods for the U.S. state-level unemployment data in terms of mean squared prediction error (MSPE) and continuous ranked probability score (CRPS), with a lead time of six months. Both metrics are calculated over all out-of-sample periods and states.}
\end{framed}
\end{varwidth}
\end{table}

Seasonally adjusted monthly unemployment data were obtained from the publicly available Federal Reserve Bank of St. Louis {\url{(http://research.stlouisfed.org/fred2}), for a period starting in January 1976 for the twelve states that make up the U.S. Census Bureau's Midwest Region \citep[][]{jones2001two}. The period from December 2008 through June 2014 was designated as the out-of-sample period to evaluate the performance of the model. This period represents the most recent unemployment cycle caused by the Great Recession that started in 2008 and provides a nonlinear time series to assess the model. Using a lead time of six months, each model was trained on data from January 1976 through May 2008. The results in Table 3 show the BAST-RNN model to be a more accurate forecast model with higher quality uncertainty measures than the three competing models. From Figure \ref{fig:Figure_4} it is clear that all four models struggle with identifying the start of the unemployment cycle in 2009, with the BAST-RNN model generally recovering to accurately predict the states with peaks later in the cycle. Across the six states displayed in Figure \ref{fig:Figure_4} (selected to represent a range of different unemployment cycles in the region), the BAST-RNN model appears to be the most accurate in terms of forecasting the decrease in the unemployment rate during the recovery that followed the 2008 recession.

\begin{figure}[H]
  \centering
    \captionsetup{font=footnotesize}
\includegraphics[width=18cm,height=10cm]{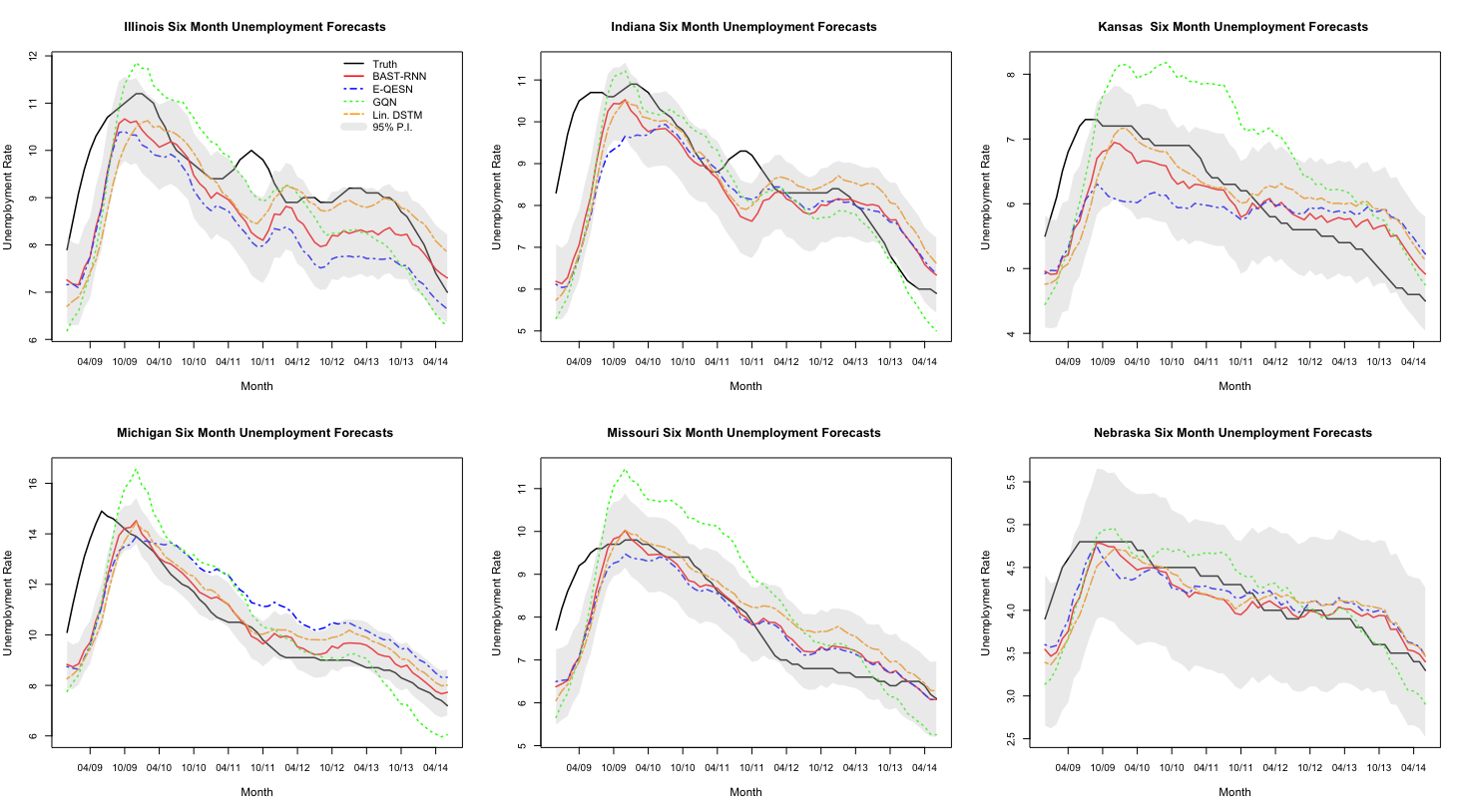} 
\caption{Posterior out-of-sample summaries and comparison methods for 6 of the 12 states in the U.S. state-level unemployment application. The observed unemployment rate is represented by the black line. The red line denotes the posterior mean from the BAST-RNN model, while the dotted blue, green, and orange line represent the E-QESN, GQN, and Linear DSTM model, respectively. The shaded grey area in each plot signifies the 95\% posterior prediction intervals associated with the BAST-RNN model.}  
\label{fig:Figure_4}
\end{figure}

\section{Discussion and Conclusion}
The results of all three applications presented above demonstrate the potential of using machine learning methods within a Bayesian modeling framework for forecasting nonlinear spatio-temporal processes. While many methods struggled with forecasting the 2015-2016 ENSO cycle, the BAST-RNN model forecasted much of the overall cycle correctly, especially when accounting for forecast uncertainty. Additionally, the BAST-RNN model was able to forecast the correct nonlinear trajectory for the multiscale Lorenz data despite a considerable amount of noise. With regards to both forecast accuracy and quantification of uncertainty, the BAST-RNN model was superior to the three competing models, over a reasonably long out-of-sample temporal span across three different applications.  

 Placing popular machine learning methods, such as RNNs, within a more rigorous statistical framework allows for more thorough uncertainty quantification, while also providing a useful framework for building more complicated models. That is, the proposed BAST-RNN model provides a first step towards more hierarchical machine learning methods that account for sources of variation at multiple levels. High-dimensional real-world processes often contain multiple layers of interconnected uncertainties and these uncertainties can more easily be untangled and formally modeled within the proposed modeling framework. Conversely, even the most precise uncertainty quantification methods are of diminishing value if they are not flexible enough to accurately forecast the process of interest. Thus, by combining the forecasting ability of the RNN model with the rigor of Bayesian modeling, the proposed methodology  provides a powerful tool for modelers.
 
 The proposed model can be used for a broad range of forecasting problems (as seen by the variety of applications analyzed here) both in its current form and with relatively minor modifications. For example, the model can easily be extended to account for different types of response data such as binary or count data. Moreover, the BAST-RNN is flexible enough to deal with varying degrees of nonlinearity, whereas past statistical nonlinear forecasting models may fail with higher levels of nonlinearity. The applications shown above provide evidence of this flexibility with the model producing accurate results for both quasilinear processes (SST and unemployment) and a highly nonlinear process (Lorenz process).

 Other extensions of the model include letting the parameters associated with the embedded input and the number of hidden units vary, which could more accurately account for the uncertainty associated with these choices. Putting the model within a fully Bayesian hierarchical framework is another possible extension.  Moreover, when adding additional hidden layers to the model it may be necessary to incorporate more computational efficient methods to improve scalability, possibly borrowing ideas from the ESN literature.  It is also likely that other forms of dimension reduction may be useful when considering the BAST-RNN model for other applications. In particular, incorporating the nonlinearity or dynamics of the process explicitly in the selected dimension reduction method could be important for some applications. For large data sets, where dimension reduction is not appropriate, it may be necessary to combine the presented computational framework with other computational methods such as Langevin dynamics \citep[i.e.,][]{welling2011bayesian}.

\section*{Acknowledgements}

This work was partially supported by the U.S. National Science Foundation (NSF) and the U.S. Census Bureau under NSF grant SES-1132031, funded through the NSF-Census Research Network (NCRN) program.
\newpage
\bibliography{reference}
\bibliographystyle{asa}
\newpage
\section*{Appendix A: Specification of Priors}
\appendix
\setcounter{equation}{0}
\renewcommand{\theequation}{A\arabic{equation}}
\begin{singlespace}
Below is a list of prior distributions for all of the parameters in the BAST-RNN model, along with specific values for each of the hyper-parameters used in the model.
\end{singlespace}
\begin{table}[H]
\captionsetup{font=footnotesize}
\renewcommand{\arraystretch}{.7}
\footnotesize
\begin{tabular}{p{15cm} p{2cm}}
 &   \\
 \hline
  &   \\
Each element in the weight matrix $\bW$ is given the following prior  distribution: \\
  &   \\
 $w_{i,\ell}=\gamma_{i,\ell}^w \text{TN}_{[-a_w,a_w]}(0,\sigma^2_{w,0})+(1-\gamma_{i,\ell}^w) \text{TN}_{[-a_w,a_w]}(0,\sigma^2_{w,1})$, for  \hspace{.15cm} $\gamma_{i,\ell}^w\sim \text{Bernoulli}(\pi_w)$, \\
  &   \\
 where $\sigma^2_{w,0}=(1,000)^2$, $\sigma^2_{w,1}=.001$, $a_w=.20$, and $\pi_w=.20$. \\
     &   \\
Each element in the weight matrix $\bU$ is given the following prior distribution: \\
 $ u_{i,r}=\gamma_{i,r}^u \text{TN}_{[-a_u,a_u]}(0,\sigma^2_{u,0})+(1-\gamma_{i,r}^u) \text{TN}_{[-a_u,a_u]}(0,\sigma^2_{u,1})$, for  \hspace{.15cm} $\gamma_{i,r}^u \sim \text{Bernoulli}(\pi_u)$, \\
  &   \\
 where $\sigma^2_{u,0}=(1,000)^2$, $\sigma^2_{u,1}=.0005$, $a_u=.20$, and $\pi_u=.025$.    \\ 
         &   \\
    Each element in the weight matrix $\bV_1$ is given the following prior distribution: \\
        &   \\
    $ v_{1,k,i}=\gamma_{1,k,i}^v \text{Gau}(0,\sigma^2_{v_1,0})+(1-\gamma_{1,k,i}^v) \text{Gau}(0,\sigma^2_{v_1,1})$, for  \hspace{.15cm} $\gamma_{1,k,i}\sim \text{Bernoulli}(\pi_{v_1})$,  \\
       &    \\
       where $\sigma^2_{v_1,0}=10, \sigma^2_{v_1,1}=.01$, and $\pi_{v_1}=.50$. \\
       &   \\
           Each element in the weight matrix $\bV_2$ is given the following prior distribution: \\
        &   \\
    $ v_{2,k,i}=\gamma_{2,k,i}^v \text{Gau}(0,\sigma^2_{v_2,0})+(1-\gamma_{2,k,i}^v) \text{Gau}(0,\sigma^2_{v_2,1})$, for  \hspace{.15cm}  $\gamma_{2,k,i}\sim \text{Bernoulli}(\pi_{v_2})$,  \\
       &    \\
       where $\sigma^2_{v_2,0}=.5, \sigma^2_{v_2,1}=.05$, and $\pi_{v_2}=.25$. \\
       &   \\
       Finally, ${\mbv \alpha} \sim\text{Gau}({\mbv 0}, \sigma^2_\alpha \bI)$, where $ \sigma_\alpha^2=(.10)^2$, \hspace{.25cm} ${\mbv \mu}\sim \text{Gau}({\mbv 0}, \sigma_\mu^2 \bI)$, where $  \sigma_\mu^2=100$, \hspace{.25cm}  $\delta\sim\text{Unif}(0,1)$, \\
          $\sigma^2_\epsilon \sim \text{IG}(\alpha_\epsilon,\beta_\epsilon)$, where $\alpha_\epsilon=.001$ and  $\beta_\epsilon=.001$. \\
                &   \\
  \hline
\end{tabular}
\end{table}

\vspace{.5cm}
\begin{singlespace}
{\small
For the SST application, $\pi_u$ was set to $.05$ for all of the $\bU$ parameters associated with the first EOF as well as $\bU$ parameters associated with non-lagged inputs corresponding to the second EOF. Overall, the first two EOFs account for almost $57\%$ of the variation in the SST data with the first EOF accounting for $46\%$ of the overall variation by itself, thus suggesting higher prior probabilities for these inputs to be included in the model. }
\end{singlespace}

\newpage

\begin{singlespace}
As discussed in the main text, the U.S. state-level unemployment application involved a much slower moving process than the SST and Lorenz system examples. In particular, this can be seen by comparing Figure \ref{fig:Figure_3} and \ref{fig:Figure_4}, where Figure \ref{fig:Figure_3} displays an approximately completed ENSO cycle, and Figure \ref{fig:Figure_4} shows part of an unemployment cycle, while Figure \ref{fig:Figure_4} covers a temporal span twice as long as Figure \ref{fig:Figure_3}. To account for this difference, the hyper-parameters $a_w$ and $a_u$ were adjusted for the unemployment application such that, $a_u=.05$ and $a_w=.05$.

To justify these prior choices we simulated from (\ref{eq:h2}) with different values of $a_w$ and $a_u$, while setting the input to zero for every period after the first period. As shown in Figure \ref{fig:Figure_5}, this procedure allowed us to examine the memory of the hidden units for different values of $a_w$ and $a_u$. The results in Figure \ref{fig:Figure_5} show that for smaller values of $a_w$ and $a_u$, the hidden units have more memory (i.e., the original signal dies off slowly). Therefore smaller values of $a_w$ and $a_u$ are more appropriate for the slower moving unemployment example, which requires more memory of the recent trajectory of the process. It is important to note that the focus of Figure \ref{fig:Figure_5} is on how quickly the orginal signal is forgotten for values of $a_w$ and $a_u$ and not the displayed dynamical speeds. 
\end{singlespace}

\vspace{.5cm}

\begin{figure}[H]
  \centering
    \captionsetup{font=footnotesize}
\includegraphics[width=10.5cm,height=8.5cm]{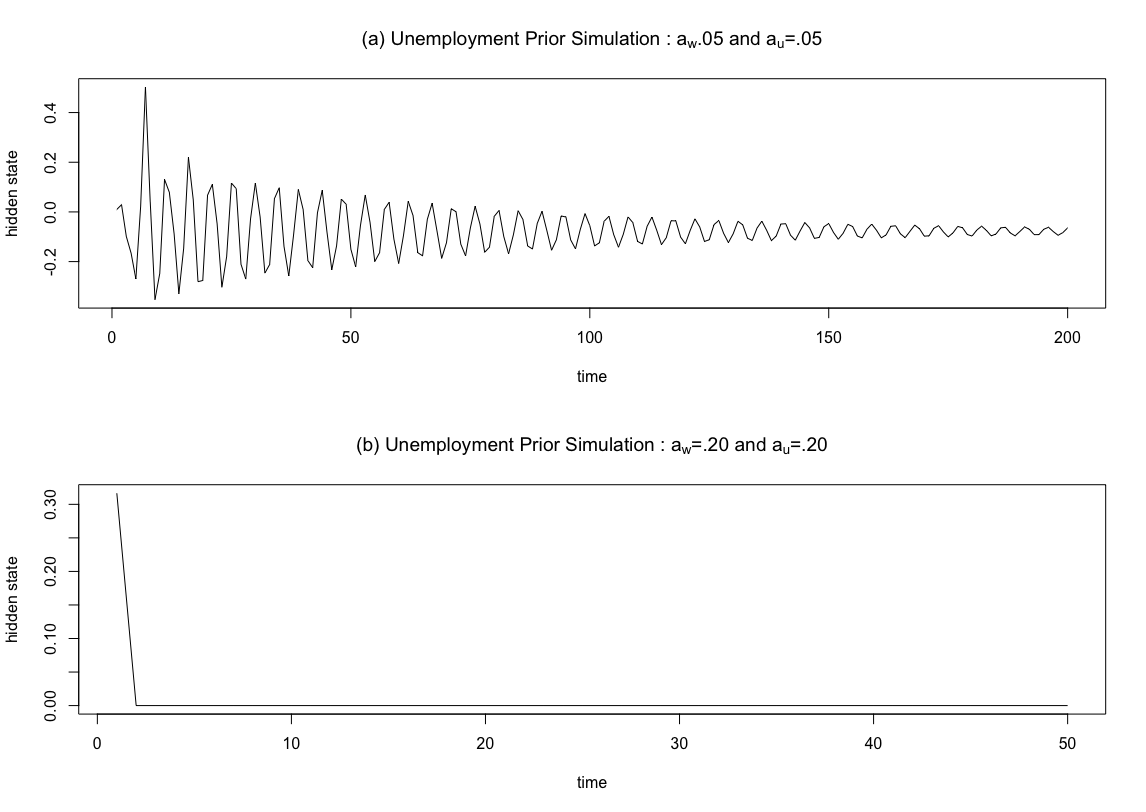} 
\caption{Prior simulation for the unemployment application to demonstrate the difference in memory of the hidden units, $\bh_t$, for different values of $a_w$ and $a_u$. (a) The first 200 periods from a simulation from (\ref{eq:h2}) with $a_w=.05$ and $a_u=.05$, for a fixed input. (b) The first 50 periods from a simulation from (\ref{eq:h2}) with $a_w=.20$ and $a_u=.20$, for a fixed input. Together, (a) and (b) illustrate how smaller values of $a_w$ and $a_u$  are more appropriate for slower moving processes that require more memory, such as the unemployment application.}  
\label{fig:Figure_5}
\end{figure}

\newpage

\section*{Appendix B: Details of Algorithm 1}

Suppose we introduce the expansion parameter ${\mbv \alpha }$  where ${\mbv \alpha}=\{ \alpha_{i,\ell}\}$  for $i=1,\dots,n_h$ and $\ell=1,\dots,n_h$, and ${\mbv \alpha} \subset \bA$ where $\bA \in \R^{n_h^2}$. Define the following function $t_{\mbv \alpha} : \bW \longrightarrow \bW$, where we require that $t_{\mbv \alpha} $ is a one-to-one differentiable function. Let $\Theta$ represent all of the parameters in the model not associated with $\bW$, such that $\Theta\equiv \{\bV_1,\bV_2, {\mbv \mu}, {\mbv\Gamma}_{V_1}, {\mbv \Gamma}_{V_2}, \bU, {\mbv \Gamma}_U, \delta, \sigma^2_\epsilon  \}$. Next, let $\bY_{1:T}\equiv \{\bY_1,\dots,\bY_T\}$ and $\widetilde{\bX}_{1:T}\equiv \{ \widetilde{\bX}_1,\dots,\widetilde{\bX}_T\}$. 

\vspace{.5cm}

We define the likelihood of the model (before the introduction of the expansion parameter matrix ${\mbv \alpha }$) using the following slight abuse of notation $\prod\limits^T_{t=1} [\bY_t \mid \Theta, \bW,{\mbv \Gamma}_W,\widetilde{\bX}_t ] = [\bY_{1:T} \mid \Theta, \bW,{\mbv \Gamma}_W,\widetilde{\bX}_{1:T} ]$, with the notation $[\cdot]$ denoting a distribution. We assume that ${\mbv \alpha }$ is only dependent on $\Theta,\bW, {\mbv \Gamma}_W,$ and $\bY_{1:T}$ through the transformation $t_{\mbv \alpha} $ and independent of these values otherwise.

\vspace{.5cm}

The function $t_{\mbv \alpha}$ is defined as follows: $t_{\mbv \alpha}(\bW) =\{t_{\alpha_{i,\ell}} (w_{i,\ell})\}=\{ \kappa( w_{i,\ell}-  \alpha_{i,\ell}) \}$, where: $\kappa(q_\kappa)=-a + \frac{2a}{1+\text{e}^{-q_\kappa}}$, thus ensuring $q_\kappa \in[-a,a] $. The Jacobian for the transformation $t_{\mbv \alpha}(\bW)$, is defined as $J_{\mbv \alpha}(\bW)=\frac{\partial}{\partial \bW} t_{\mbv \alpha} (\bW)= \prod\limits_{i=1}^{n_h} \prod\limits_{\ell=1}^{n_h}  \frac{\partial t_{\alpha_{i,\ell}} ( w_{i,\ell})}{\partial w_{i,\ell}}  $, while the Jacobian for the transformation $t^{-1}_{\mbv \alpha}(\bW)$, is defined as $\widetilde{J}_{\mbv \alpha}(\bW)=\frac{\partial}{\partial \bW} t^{-1}_{\mbv \alpha} (\bW)= \prod\limits_{i=1}^{n_h} \prod\limits_{\ell=1}^{n_h}  \frac{\partial t^{-1}_{\alpha_{i,\ell}} ( w_{i,\ell})}{\partial w_{i,\ell}}  $.

{\bf Details for Algorithm 1}

1. Sample $\bW$ and ${\mbv \Gamma}_W$ as follows:
 
\vspace{-1.35cm}

\begin{align}
[\bW,{\mbv \Gamma}_W \mid \Theta,\bY_{1:T}, \widetilde{\bX}_{1:T}  ] & =\frac{ [\Theta,\bW,{\mbv \Gamma}_W \mid \bY_{1:T}, \widetilde{\bX}_{1:T}  ]}{[\Theta \mid \bY_{1:T}, \widetilde{\bX}_{1:T}  ]} \label{A1}   \\
&=\frac{\int_\bA  [\Theta,\bW,{\mbv \Gamma}_W,{\mbv \alpha} \mid \bY_{1:T}, \widetilde{\bX}_{1:T}  ]d{\mbv \alpha} }{[\Theta \mid \bY_{1:T}, \widetilde{\bX}_{1:T}  ]}  \label{A2}  \\
&=\frac{\int_\bA  [\Theta,\bW,{\mbv \Gamma}_W\mid \bY_{1:T}, \widetilde{\bX}_{1:T},{\mbv \alpha}   ]  [{\mbv \alpha}  \mid \bY_{1:T}, \widetilde{\bX}_{1:T}] d{\mbv \alpha} }{[\Theta \mid \bY_{1:T}, \widetilde{\bX}_{1:T},  ]} \label{A3}   
\end{align}
\begin{align}
&=\frac{\int_\bA  [\Theta,\bW,{\mbv \Gamma}_W\mid \bY_{1:T}, \widetilde{\bX}_{1:T} ]  [{\mbv \alpha} ] d{\mbv \alpha} }{[\Theta \mid \bY_{1:T}, \widetilde{\bX}_{1:T},  ]}  \label{A4}   \\
& =\frac{\int_\bA  [\Theta,t_{\mbv \alpha}(\bW),{\mbv \Gamma}_W \mid \bY_{1:T}, \widetilde{\bX}_{1:T}  ] \ | J_{\mbv \alpha}(\bW)| \ [{\mbv \alpha}] \ d{\mbv \alpha} }{[\Theta \mid \bY_{1:T}, \widetilde{\bX}_{1:T}  ]} \label{A5}   \\
& =  \int_\bA  [t_{\mbv \alpha}(\bW),{\mbv \Gamma}_W\mid \Theta,\bY_{1:T}, \widetilde{\bX}_{1:T}  ] \ | J_{\mbv \alpha}(\bW)| \ [{\mbv \alpha}] \ d{\mbv \alpha},  \label{A6}  
\end{align}

     As stated in the main text, to sample from this integral, we assume $\bW', {\mbv \Gamma}_W \sim [t_{\mbv \alpha}(\bW),{\mbv \Gamma}_W\mid \Theta,\bY_{1:T},\widetilde{\bX}_{1:T}]$. We will take $\bW=t_{{\mbv \alpha}}^{-1}(\bW')$, thus allowing for the joint sampling of $\bW$ and ${\mbv \Gamma}_W$. This result leads to step 1 of Algorithm 1 and defining ${\mbv \alpha}_0$ as the simulated value from $[{\mbv \alpha}]$ leads to step 2. Note the procedure described here closely follows the procedure outlined directly below equation (1.4.3) in \cite{hobert2008theoretical}. The assumption stated above that ${\mbv \alpha }$ is only dependent on $\Theta,\bW, {\mbv \Gamma}_W,\widetilde{\bX}_{1:T} $ and $\bY_{1:T}$ through the transformation $t_{\mbv \alpha} $ is utilized when going from (\ref{A3}) to (\ref{A4}).

       \vspace{.05cm} 
   
        2. Sample $\Theta$ and ${\mbv \alpha}$, as follows:
           
             \vspace{-1.575cm}

 \begin{align}
[\Theta,{\mbv \alpha} \mid \bW,{\mbv \Gamma}_W,\bY_{1:T}, \widetilde{\bX}_{1:T}   ] &=  \frac{[\Theta,{\mbv \alpha}, \bW,{\mbv \Gamma}_W,\bY_{1:T}, \widetilde{\bX}_{1:T}   ] }{ [\bW,{\mbv \Gamma}_W,\bY_{1:T}, \widetilde{\bX}_{1:T} ]}  \label{A7}   \\
&=  \frac{[\Theta,{\mbv \alpha}, t^{-1}_{{\mbv \alpha}_0}(\bW),{\mbv \Gamma}_W,\bY_{1:T}, \widetilde{\bX}_{1:T}   ]  \ | \widetilde{J}_{{\mbv \alpha}_0}(\bW) | }{ [t^{-1}_{{\mbv \alpha}_0}(\bW),{\mbv \Gamma}_W,\bY_{1:T}, \widetilde{\bX}_{1:T}   ]  \ | \widetilde{J}_{{\mbv \alpha}_0}(\bW) |} \label{A8}  \\
&=  \frac{[\Theta,{\mbv \alpha}, t^{-1}_{{\mbv \alpha}_0}(\bW),{\mbv \Gamma}_W,\bY_{1:T}, \widetilde{\bX}_{1:T}   ] }{ [t^{-1}_{{\mbv \alpha}_0}(\bW),{\mbv \Gamma}_W,\bY_{1:T}, \widetilde{\bX}_{1:T}   ]} \label{A9}   \\ 
& \propto  [\Theta,{\mbv \alpha},\widetilde{\bW },{\mbv \Gamma}_W,\bY_{1:T}, \widetilde{\bX}_{1:T} ] \label{A10}   \\
&= [\Theta,{\mbv \alpha}, t_{\mbv \alpha} (\widetilde{\bW }),{\mbv \Gamma}_W,\bY_{1:T}, \widetilde{\bX}_{1:T}   ]  \ |  J_{\mbv \alpha} (\widetilde{\bW }) | \label{A11}   \\
& \propto  [\bY_{1:T} \mid  t_{\mbv \alpha} (\widetilde{\bW }),{\mbv \Gamma}_W, \Theta,{\mbv \alpha} , \widetilde{\bX}_{1:T}  ] \ [\Theta] \nonumber \\ 
&  \times [  t_{\mbv \alpha} (\widetilde{\bW }) \mid {\mbv \Gamma}_W, {\mbv \alpha}   ] \ [{\mbv \alpha} ]  \ |  J_{\mbv \alpha} (\widetilde{\bW }) | \label{A12}  
\end{align}
             
          \vspace{.25cm} 
            \noindent    Above in (\ref{A10}), $\widetilde{\bW }$ is defined as $\widetilde{\bW }\equiv t^{-1}_{{\mbv \alpha}_0}(\bW)$. Going from (\ref{A11}) to (\ref{A12}), we assume ${\mbv \alpha} $ is conditionally independent of $ {\mbv \Gamma}_W$, $t_{\mbv \alpha} (\widetilde{\bW })$ and  ${\mbv \alpha} $ are independent of $\widetilde{\bX}_{1:T} $, and $\Theta$ is conditionally independent of ${\mbv \alpha},t_{\mbv \alpha} (\widetilde{\bW }) , {\mbv \Gamma}_W,$ and $\widetilde{\bX}_{1:T} $. Finally, the full-conditional distributions for $\Theta$ and ${\mbv \alpha}$ are as follows: 
       \begin{align}
[ {\mbv \alpha} \mid  t_{\mbv \alpha} (\widetilde{\bW }),{\mbv \Gamma}_W,\Theta, \bY_{1:T},, \widetilde{\bX}_{1:T}  ] & \propto [\bY_{1:T} \mid  t_{\mbv \alpha} (\widetilde{\bW }),{\mbv \Gamma}_W, \Theta,{\mbv \alpha},\widetilde{\bX}_{1:T}  ] \nonumber  \\ 
& \times  [  t_{\mbv \alpha} (\widetilde{\bW }) \mid {\mbv \Gamma}_W, {\mbv \alpha}   ] \ [{\mbv \alpha} ]  \ |  J_{\mbv \alpha} (\widetilde{\bW }) |   \label{A13}  
                \end{align}
               
                              \vspace{-1.75cm} 

                \begin{align}
                [ \Theta \mid  t_{\mbv \alpha} (\widetilde{\bW }),{\mbv \Gamma}_W,{\mbv \alpha}, \bY_{1:T}, \widetilde{\bX}_{1:T}  ] & \propto [\bY_{1:T} \mid  t_{\mbv \alpha} (\widetilde{\bW }),{\mbv \Gamma}_W, \Theta,{\mbv \alpha},\widetilde{\bX}_{1:T}  ] \ [\Theta] \label{A14}   \\
               & \propto   [\bY_{1:T} \mid  t_{\mbv \alpha} (\widetilde{\bW }), \Theta,\widetilde{\bX}_{1:T}  ] \ [\Theta]  \label{A15}  
                \end{align}

                   \noindent        These two full conditionals lead directly to steps 4 and 5 of Algorithm 1, respectively, where ${\mbv \alpha}$ is sampled using Metropolis-Hasting steps and $\Theta$ is sampled using Gibbs and Metropolis-Hasting steps (see Appendix C).

      \vspace{.85cm} 
      
    \setcounter{algorithm}{0}
\begin{algorithm}[H]
\singlespace
\begin{enumerate}
\item Sample $\bW,{\mbv \Gamma}_W$ from: $ [\bW,{\mbv \Gamma}_W\mid \Theta,\bY_{1:T},\widetilde{\bX}_{1:T} ] \propto$ 

\vspace{1mm} 

$[\bY_{1:T}\mid  \Theta, \bW,{\mbv \Gamma}_W, \widetilde{\bX}_{1:T}  ] \ [\bW\mid {\mbv \Gamma}_W] \ [{\mbv \Gamma}_W]$.
\item Generate $\alpha_{0,i,\ell}\sim \text{Gau}(0, \sigma^2_\alpha )$ for $i=1,\dots,n_h$ and $\ell=1,\dots,n_h$. 
\item Transform $\widetilde{\bW}=t^{-1}_{{\mbv \alpha}_0}(\bW)$.
\item Sample ${\mbv \alpha}$ from: $[ {\mbv \alpha} \mid  t_{\mbv \alpha} (\widetilde{\bW }),{\mbv \Gamma}_W,\Theta, \bY_{1:T} ,\widetilde{\bX}_{1:T}] \propto $

\vspace{1mm}

$[\bY_{1:T} \mid  t_{\mbv \alpha} (\widetilde{\bW }),{\mbv \Gamma}_W, \Theta,{\mbv \alpha},\widetilde{\bX}_{1:T}  ] \  [  t_{\mbv \alpha} (\widetilde{\bW }) \mid {\mbv \Gamma}_W, {\mbv \alpha}   ] \ [{\mbv \alpha} ]  \ |  J_{\mbv \alpha} (\widetilde{\bW }) | $.
\item Sample $\Theta$ from: $ [ \Theta \mid  t_{\mbv \alpha} (\widetilde{\bW }),{\mbv \Gamma}_W,{\mbv \alpha}, \bY_{1:T}, \widetilde{\bX}_{1:T}  ] \propto   [\bY_{1:T} \mid  t_{\mbv \alpha} (\widetilde{\bW }), \Theta,\widetilde{\bX}_{1:T}  ] \ [\Theta].$
\end{enumerate}

\vspace{3mm}

\caption{PX-MCMC algorithm}

\end{algorithm}

\newpage

\section*{Appendix C: Full-Conditionals for the BAST-RNN model}
The full-conditional distributions for all of the parameters in the BAST-RNN model are detailed in this Appendix. To ease the notation we define:
\begin{eqnarray}
 \widetilde{\Theta}\equiv \{{\mbv \mu}, \bV_1,\bV_2, {\mbv\Gamma}_{V_1}, {\mbv \Gamma}_{V_2}, \bW, {\mbv \Gamma}_W, {\mbv \alpha},{\mbv \alpha}_0 ,\bU, {\mbv \Gamma}_U, \delta, \sigma^2_\epsilon \}, \nonumber
\end{eqnarray} 
and borrowing the notational convention from \cite{bradley2016bayesian}, let \\ $\widetilde{\Theta}_{-w_{i,\ell}}=\widetilde{\Theta}\ \cap \{ w_{i,\ell}\}^c$, such that the notation ``c" denotes the compliment. Thus, $\widetilde{\Theta}_{-w_{i,\ell}}$ denotes the collection of all of the parameters in $\widetilde{\Theta}$ except for $w_{i,\ell}$. A similar notation can be used for all of the other parameters in the model. 

We will use the notation $\Phi(\cdot)$ to denote the cumulative distribution function for the Gaussian distribution. Next, let $\Upsilon_k$ be a $(2n_h+1) \times (2n_h+1)$ diagonal matrix with the first diagonal element corresponding to $\sigma^2_{\mu}$, the next $n_h$ diagonal entries corresponding to $\gamma_{1,k,i}^v \sigma^2_{v_1,0}+(1-\gamma_{1,k,i}^v) \sigma^2_{v_1,1}$ (for $i=1,\dots,n_h$), and the last $n_h$ diagonal entries corresponding to $\gamma_{2,k,i}^v\sigma^2_{v_2,0}+(1-\gamma_{2,k,i}^v) \sigma^2_{v_2,1}$ (for $i=1,\dots,n_h$). Probability distribution functions for the Gaussian priors associated with $v_{1,k,i}$ and $v_{2,k,i}$ are denoted by $\phi^{v_1}(\cdot)$ and  $\phi^{v_2}(\cdot)$ (as defined in Appendix A), respectively.

Finally, the vector $\widetilde{\bh}_t$ is defined as $\widetilde{\bh}_t\equiv(1,\bh_t',\bh_t^{2'})'$, such that $\widetilde{\bh}_{1:T}$ is a $(2n_h+1) \times T$ matrix. Throughout, we will let $\bg_t \equiv {\mbv \mu}+  \bV_1\bh_t + \bV_2\bh_t^2$ to reduce the amount of notation. The BAST-RNN model is defined by the following full-conditional distributions:

      \vspace{.5cm} 
      
      \begin{itemize}
      
     \item $[w_{i,\ell},\gamma_{i,\ell}^w \mid \bY_{1:T}, \widetilde{\bX}_{1:T}  , \widetilde{\Theta}_{-\{w_{i,\ell}, \gamma_{i,\ell}^w \} } ] \propto  \prod\limits^T_{t=1} \text{exp} \bigg( \frac{ -( \bY_t- \bg_t  )'  (\bY_t- \bg_t  ) }{2\sigma^2_\epsilon} \bigg) \\ \times \bigg(  \frac{\gamma_{i,\ell}^w \text{exp} \big( \frac{-w_{i,\ell}}{2 \sigma^2_{w,0}}  \big) }{\Phi(\frac{a_w}{\sigma_{w,0}}) - \Phi(\frac{-a_w}{\sigma_{w,0}})}  + \frac{(1-\gamma_{i,\ell}^w) \text{exp} \big( \frac{-w_{i,\ell}}{2 \sigma^2_{w,1}}  \big) }{\Phi(\frac{a_w}{\sigma_{w,1}}) - \Phi(\frac{-a_w}{\sigma_{w,1}})} \bigg) $ $\times  \bigg( \pi_w^{\gamma_{i,\ell}^w} + (1-\pi_w)^{1- \gamma_{i,\ell}^w} \bigg) $,
     
          \vspace{.25cm} 
          
          for $i=1,\dots,n_h$  and $\ell=1,\dots,n_h$.
          
          \item $[\alpha_{i,\ell} \mid \bY_{1:T}, \widetilde{\bX}_{1:T}  , \widetilde{\Theta}_{-\alpha_{i,\ell}}  ] \propto  \prod\limits^T_{t=1} \text{exp} \bigg( \frac{ -( \bY_t- \bg_t  )'  (\bY_t- \bg_t  ) }{2\sigma^2_\epsilon} \bigg) \\ \times \bigg(  \frac{\gamma_{i,\ell}^w \text{exp} \big( \frac{-t_{\alpha_{i,\ell}}(\tilde{w}_{i,\ell})}{2 \sigma^2_{w,0}}  \big) }{\Phi(\frac{a_w}{\sigma_{w,0}}) - \Phi(\frac{-a_w}{\sigma_{w,0}})}  + \frac{(1-\gamma_{i,\ell}^w) \text{exp} \big( \frac{-t_{\alpha_{i,\ell}}(\tilde{w}_{i,\ell})}{2 \sigma^2_{w,1}}  \big) }{\Phi(\frac{a_w}{\sigma_{w,1}}) - \Phi(\frac{-a_w}{\sigma_{w,1}})} \bigg) \times \text{exp}\bigg(\frac{-\alpha_{i,\ell}}{2\sigma^2_\alpha}\bigg) \times \bigg( \frac{2a_w \text{exp}(-\tilde{w}_{i,\ell}+\alpha_{i,\ell}) }{(1+\text{exp}(-\tilde{w}_{i,\ell}+\alpha_{i,\ell}))^2}  \bigg) $,
          
                \vspace{.25cm} 
          
          for $i=1,\dots,n_h$  and $\ell=1,\dots,n_h$.
          
               \item $[u_{i,r},\gamma_{i,r}^u \mid \bY_{1:T}, \widetilde{\bX}_{1:T}  , \widetilde{\Theta}_{-\{u_{i,r}, \gamma_{i,r}^u \} } ] \propto  \prod\limits^T_{t=1} \text{exp} \bigg( \frac{ -( \bY_t- \bg_t  )'  (\bY_t- \bg_t  ) }{2\sigma^2_\epsilon} \bigg) \\ \times \bigg(  \frac{\gamma_{i,r}^u \text{exp} \big( \frac{-u_{i,r}}{2 \sigma^2_{u,0}}  \big) }{\Phi(\frac{a_u}{\sigma_{u,0}}) - \Phi(\frac{-a_u}{\sigma_{u,0}})}  + \frac{(1-\gamma_{i,r}^u) \text{exp} \big( \frac{-u_{i,r}}{2 \sigma^2_{u,1}}  \big) }{\Phi(\frac{a_u}{\sigma_{u,1}}) - \Phi(\frac{-a_u}{\sigma_{u,1}})} \bigg) $ $\times  \bigg( \pi_u^{\gamma_{i,r}^u} + (1-\pi_u)^{1- \gamma_{i,r}^u} \bigg) $,
     
          \vspace{.25cm} 
          
          for $i=1,\dots,n_h$  and $\ell=1,\dots,n_h$.

            \vspace{.25cm} 
            
                 \item $[\delta \mid \bY_{1:T}, \widetilde{\bX}_{1:T}  , \widetilde{\Theta}_{-\delta}  ] \propto  \prod\limits^T_{t=1} \text{exp} \bigg( \frac{ -( \bY_t- \bg_t  )'  (\bY_t- \bg_t  ) }{2\sigma^2_\epsilon} \bigg) \times  I_{[0,1]}(\delta)$
            
 \vspace{.25cm} 
 
 \item $[\mu_{1,k},\bV_{1,k}, \bV_{2,k} \mid \bY_{1:T}, \widetilde{\bX}_{1:T}  , \widetilde{\Theta}_{-\{\mu_{1,k},\bV_{1,k}, \bV_{2,k} \} } ] \propto \\ \text{Gau} \bigg( \big( \frac{1}{\sigma^2\epsilon} \widetilde{\bh}_{1:T} \widetilde{\bh}_{1:T} ' + \Upsilon_k^{-1}\big)^{-1}   \frac{1}{\sigma^2\epsilon}\widetilde{\bh}_{1:T} \bY_{1:T,k} ,  \big( \frac{1}{\sigma^2\epsilon} \widetilde{\bh}_{1:T} \widetilde{\bh}_{1:T} ' + \Upsilon_k^{-1}\big)^{-1}   \ \bigg)$,
 
      for $k=1,\dots,n_h$.

            \vspace{.25cm} 
            
            \item $[\gamma_{1,k,i}^v \mid  \bY_{1:T}, \widetilde{\bX}_{1:T}  , \widetilde{\Theta}_{-\{ \gamma_{1,k,i}^v \} } ] \propto \text{Bernoulli}  \bigg(\frac{\phi^{v_1}(v_{1,k,i} \mid \gamma_{1,k,i}^v=1) }{\phi^{v_1}(v_{1,k,i} \mid \gamma_{1,k,i}^v=1) + \phi^{v_1}(v_{1,k,i} \mid \gamma_{1,k,i}^v=0)}   \bigg)  $,
            
                      for $i=1,\dots,n_h$  and $k=1,\dots,n_y$.
                      
                                  \vspace{.25cm} 
            
            \item $[\gamma_{2,k,i}^v \mid  \bY_{1:T}, \widetilde{\bX}_{1:T}  , \widetilde{\Theta}_{-\{ \gamma_{2,k,i}^v \} } ] \propto \text{Bernoulli}  \bigg(\frac{\phi^{v_2}(v_{2,k,i} \mid \gamma_{2,k,i}^v=1) }{\phi^{v_2}(v_{2,k,i} \mid \gamma_{2,k,i}^v=1) + \phi^{v_2}(v_{2,k,i} \mid \gamma_{2,k,i}^v=0)}   \bigg)  $,
            
                      for $i=1,\dots,n_h$  and $k=1,\dots,n_y$.
            
               \vspace{.25cm} 
            
             \item $[\sigma^2_\epsilon \mid \bY_{1:T}, \widetilde{\bX}_{1:T}  , \widetilde{\Theta}_{-\{ \sigma^2_\epsilon \} } ] \propto \text{IG}(\frac{Tn_y}{2}+\alpha_\epsilon,\frac{1}{2}\sum\limits_{t=1}^T ( \bY_t- \bg_t  )'  (\bY_t- \bg_t  )+\beta_\epsilon)$.

\end{itemize}

All the parameters in $\widetilde{\Theta}\ $ are sampled in the order provided above, with $\bW, {\mbv \Gamma}_W, {\mbv \alpha}, \bU, {\mbv \Gamma}_U,$ and $\delta$ requiring Metropolis-Hasting steps, while $\bV_1,\bV_2, {\mbv \mu}, {\mbv\Gamma}_{V_1}, {\mbv \Gamma}_{V_2}$, and $\sigma^2_\epsilon$ are sampled with Gibbs steps.

\newpage

\section*{Appendix D: Trace Plots for the BAST-RNN Model}
\begin{singlespace}
Displayed below are various trace plots from the Lorenz-96 simulated example. The convergence for the other applications (not shown here) was similar. 
\end{singlespace}
\begin{figure}[H]
  \centering
    \captionsetup{font=footnotesize}
\includegraphics[width=12cm,height=8cm]{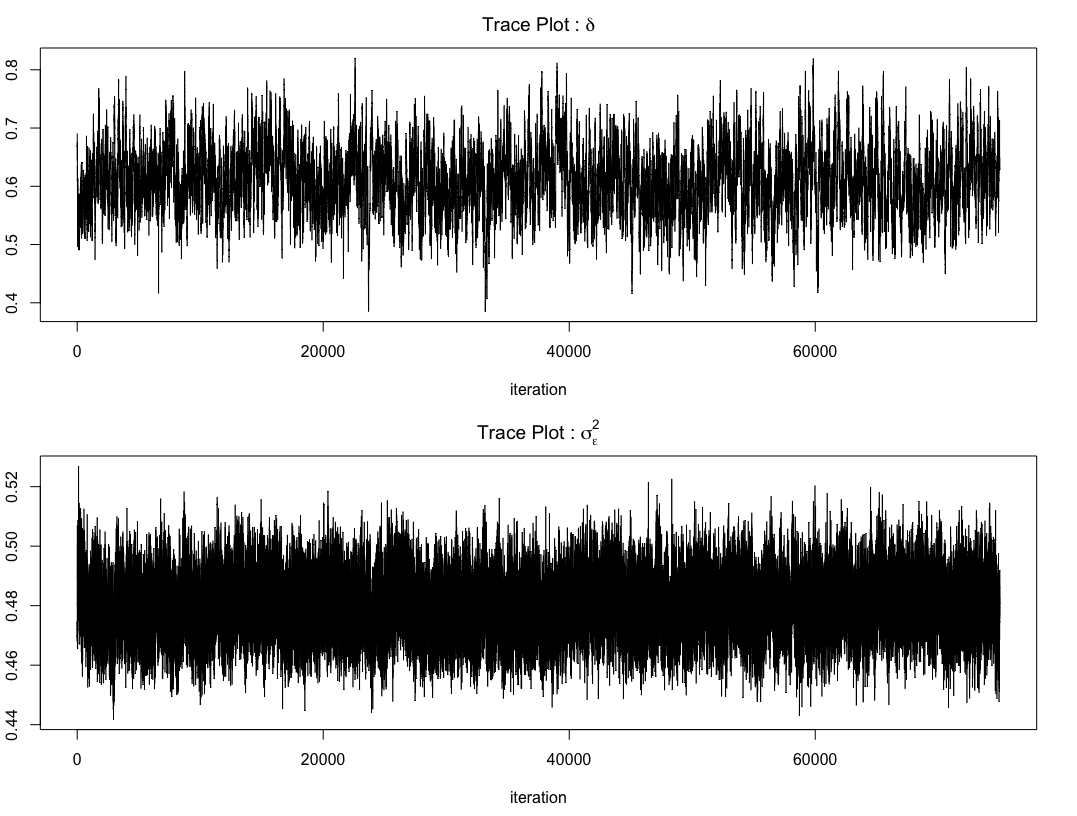}
\caption{Trace plots for the parameters $\delta$ and $\sigma^2_\epsilon$ for the Lorenz-96 simulated example.}  
\label{fig:Figure_6}
\end{figure}

\vspace{.5cm}

\begin{figure}[H]
  \centering
    \captionsetup{font=footnotesize}
\includegraphics[width=12cm,height=8cm]{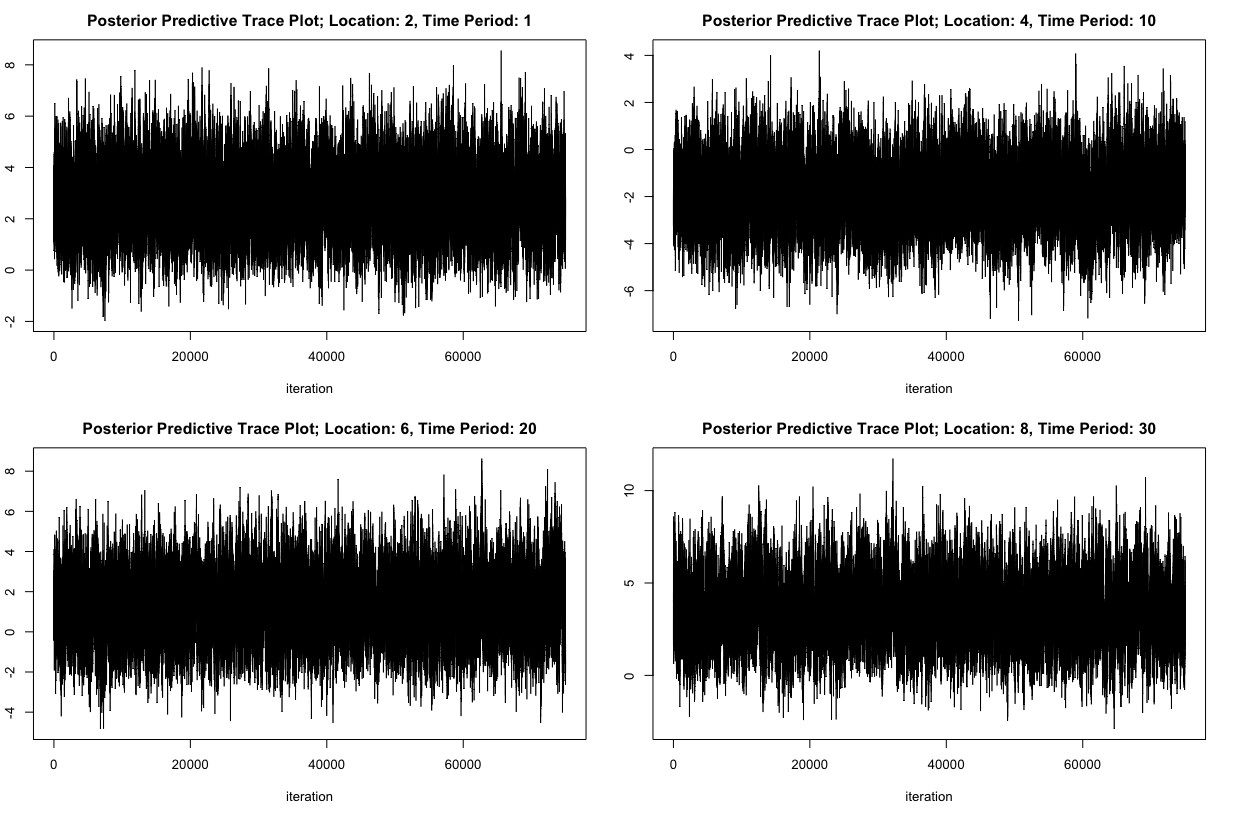}
\caption{A sample of four trace plots for the posterior predictions from the Lorenz-96 simulated example.}  
\label{fig:Figure_7}
\end{figure}

\end{document}